\documentclass[10pt,aps,prd,twocolumn,notitlepage,nofootinbib,floatfix,superscriptaddress]{revtex4-1}

\usepackage{amsmath, amsthm, amssymb, amsfonts, amsbsy, mathrsfs}

\usepackage{xcolor}
\usepackage{calligra,bm}
\usepackage{stmaryrd}

\usepackage{caption}
\usepackage{subcaption}
\usepackage{graphicx}
\graphicspath{ {./figures/} }
\usepackage{float}

\usepackage[hidelinks,bookmarks=true]{hyperref} 
\hypersetup{pdfstartview=FitH,pdfhighlight=/O,colorlinks=false}

\begin{document}

\title{Introducing a multiscale feature integration network for inpainting with applications to enhanced CMB map reconstruction}
\author{Reyhan D. Lambaga}
\email{d09244004@ntu.edu.tw}
\affiliation{Graduate Institute of Astrophysics, National Taiwan University, Taipei 10617, Taiwan}
\affiliation{Department of Physics and Center for Theoretical Sciences, National Taiwan University, Taipei 10617, Taiwan}
\affiliation{Leung Center for Cosmology and Particle Astrophysics, National Taiwan University, Taipei 10617, Taiwan}
\author{Vipin Sudevan}
\email{vipinsudevan1988@gmail.com}
\affiliation{Leung Center for Cosmology and Particle Astrophysics, National Taiwan University, Taipei 10617, Taiwan}
\author{Pisin Chen}
\affiliation{Graduate Institute of Astrophysics, National Taiwan University, Taipei 10617, Taiwan}
\affiliation{Department of Physics and Center for Theoretical Sciences, National Taiwan University, Taipei 10617, Taiwan}
\affiliation{Leung Center for Cosmology and Particle Astrophysics, National Taiwan University, Taipei 10617, Taiwan}
\affiliation{Kavli Institute for Particle Astrophysics and Cosmology, SLAC National Accelerator Laboratory, Stanford University, Stanford, California 94305, USA}
\date{\today}

\begin{abstract}
We introduce a novel neural network, SkyReconNet, which combines the expanded receptive fields of dilated convolutional layers along with standard convolutions, to capture both the global and local features for reconstructing the missing information in an image. We implement our network to inpaint the masked regions in a full-sky Cosmic Microwave Background (CMB) map. Inpainting CMB maps is a particularly formidable challenge when dealing with extensive and irregular masks, such as galactic masks which can obscure substantial fractions of the sky. The hybrid design of SkyReconNet leverages the strengths of standard and dilated convolutions to accurately predict CMB fluctuations in the masked regions, by effectively utilizing the information from surrounding unmasked areas. During training, the network optimizes its weights by minimizing a composite loss function that combines the Structural Similarity Index Measure (SSIM) and mean squared error (MSE). SSIM preserves the essential structural features of the CMB, ensuring an accurate and coherent reconstruction of the missing CMB fluctuations, while MSE minimizes the pixel-wise deviations, enhancing the overall accuracy of the predictions. The predicted CMB maps and their corresponding angular power spectra align closely with the targets, achieving the performance limited only by the fundamental uncertainty of cosmic variance. The network’s generic architecture enables application to other physics-based challenges involving data with missing or defective pixels, systematic artefacts etc. Our results demonstrate its effectiveness in addressing the challenges posed by large irregular masks, offering a significant inpainting tool not only for CMB analyses but also for image-based experiments across disciplines where such data imperfections are prevalent.

\end{abstract}

\maketitle


\section{Introduction}
The discovery of fluctuations in the Cosmic Microwave Background (CMB) marks a monumental achievement in 
modern precision cosmology, providing a glimpse into the universe as it existed approximately 380,000 years 
after the Big Bang. Over the past decades, several groundbreaking missions such as WMAP~\citep{WMAP:2003ivt} 
and Planck~\citep{2011A&A...536A...1P} have revolutionized our understanding of the CMB, 
offering unprecedented precision in analyzing its temperature and polarization anisotropies. 
These subtle fluctuations in the temperature 
and polarization fields in CMB serve as a cosmic time capsule, revealing insights into the physics 
of the Big Bang~\citep{Durrer:2015lza} and shedding light on the energy scales associated with 
cosmic inflation~\citep{Abazajian:2013vfg,Planck:2013jfk}. 
Moreover, CMB also provides stringent 
constraints to various fundamental cosmological parameters~\cite{aghanim2020planck}, including  
neutrino masses, reionization scale, etc. 
To further improve the quality of CMB signal, numerous scientific missions~\citep{masi2002boomerang, bennett2003microwave, hincks2010atacama, gandilo2016primordial,  
gualtieri2018spider, li2019probing, hui2018bicep, abazajian2022snowmass, adak2022b, litebird2023probing} are 
in the ongoing or advanced planning stages, aiming to observe the sky with even greater precision. 
However, despite the advancements, the CMB signal observed by these scientific missions are inevitably 
contaminated  by various sources, 
such as emissions from galactic and extragalactic astrophysical sources in the 
microwave region of the frequency spectrum, the foregrounds, and the unavoidable  
instrumental noise introduced by the detectors. 

Over the years, several sophisticated methods have been developed to address these contaminations in the 
observed CMB maps. Broadly categorized into 
parametric~\citep{eriksen2007joint,eriksen2008joint,land2006template,jaffe2006fast} 
and non-parametric methods~\citep{bennett2003first,eriksen2004foreground,tegmark2003high,delabrouille2009full,sudevan2017improved,Sudevan:2018qyj,Sudevan:2017una,Taylor:2006otn,hurier2013milca}, these techniques  
aim to mitigate the adverse effects 
posed by foreground and noise contaminations in CMB observations. Parametric methods rely on 
explicit modelling of CMB,
foregrounds, etc while non-parametric methods leverages statistical features of CMB and/or foregrounds 
in order to remove them. 
But irrespective of which method is used,  the cleaned CMB maps invariably retain 
some level of residual contaminations.

To minimize the impact of such foreground/noise residuals from  introducing biases in subsequent CMB 
cosmological analyses, 
it is a common practice to mask regions in the sky exhibiting high levels of residual contaminations, 
particularly in the
vicinity of the  galactic plane and strong point sources. 
This masking strategy enables us to focus only on the regions in the sky where the signal is as clean as 
possible. 
However, for some analyses like harmonic space analysis requires full-sky maps. 
Even with masking, full-sky analyses remain feasible using the partial-sky maps, by employing 
techniques designed to either reconstruct the missing sky fractions, known as inpainting, or by using 
analytical methods like pseudo-$C_\ell$ estimators (MASTER algorithm~\citep{hivon2002master,alonso2019unified}), 
Quadratic Maximum Likelihood (QML) estimators~\citep{Tegmark:1996qt,Tegmark:2001zv,Bilbao-Ahedo:2021jhn}, etc., to estimate full-sky power spectrum.  While these analytical methods can recover 
full-sky power spectrum from a partial-sky CMB map, depending on mask area they might require higher 
apodization or larger binning in order to prevent any unphysical power at any multipoles.
By contrast, inpainting enables full-sky analyses directly from a reconstructed map without additional apodization and binning procedures, thereby preserving more sky signal.
It provides a statistical reconstruction of the CMB fluctuations in masked regions by utilizing 
the information available from the unmasked portions of the sky and the underlying statistical 
properties of the CMB. 
An inpainting algorithm on the sphere based on a sparse representation of the CMB in the spherical 
harmonics domain is developed by~\citep{ABRIAL2008289}.
This sparsity-based inpainting techniques are applied in CMB weak-lensing using Planck-simulated data~\citep{2010A&A...519A...4P,2012A&A...544A..27P}, integrated Sachs-Wolfe (ISW) effect on WMAP data~\citep{2011A&A...534A..51D}.
In both studies, the robustness of the sparsity-based inpainting approach was validated through extensive 
Monte Carlo simulations.  
Some of the analytical methods 
to inpaint CMB map involves filling  
the missing region by generating a random Gaussian realization~\cite{hoffman1991constrained,bucher2012filling,2012ApJ...750L...9K}
from a prior distribution while keeping the unmasked region as a constraint. These methods have a 
limitation of not being
suitable for non-Gaussianity studies.  A maximum-likelihood 
estimator (MLE) based approach is discussed in~\cite{2011PhRvD..84j3002F,copi2011bias} with appropriate 
assumptions such as Gaussianity. 
While \citep{Planck:2013wtn, Gimeno-Amo:2024hca,Marcos-Caballero:2019jqj} explores inpainting of CMB map by interpolating the missing pixels using Gaussian process regression, they use the best-fit power spectra to determine pixel correlations and sample missing pixels 
from the conditional probability distribution with the assumption of Gaussianity.

In recent years, with the significant progress in the field of  Machine Learning (ML), 
ML-based techniques~\citep{hochreiter1997long,goodfellow2014generative,vaswani2017attention} are 
implemented to solve various complex problems across diverse fields. For tasks involving the processing of data which has a spatial or grid-like structure, deep learning architectures like 
Convolutional Neural Networks (CNN)~\citep{krizhevsky2017imagenet} are extensively utilized. 
CNN excels in image processing tasks such as image classification~\citep{dai2021coatnet,cai2022reversible}, 
object detection~\citep{zhou2021probabilistic,maaz2022edgenext},  segmentation\citep{cai2022reversible,dai2017deformable},  inpainting~\citep{NIPS2012_6cdd60ea,pmlr-v37-sohl-dickstein15, NIPS2015_daca4121,blindinpaint,2016arXiv160407379P} etc
making them a cornerstone in the field of computer vision. The ability of CNNs to learn from data without 
explicit feature engineering allows for more efficient and potentially more accurate analysis of the complex 
problems including CMB analyses.

In the field of CMB research, neural network architectures based on CNNs have been explored extensively to 
deal with various challenges such as minimizing foreground 
contaminations in observed CMB data~\cite{petroff2020full, wang2022recovering, casas2022cenn, yan2023recovering, yan2024cmbfscnn,Sudevan:2024hwq}, 
delensing on CMB polarization maps~\cite{yan2023delensing}, inpainting~\citep{puglisi2020inpainting,yi2020cosmo,Sadr:2020rje,montefalcone2021inpainting}.
Additionally, more general ML methods have been applied in CMB data analysis, such as estimating cosmological parameters~\citep{fluri2018cosmological,fluri2019cosmological}, 
identifying foreground models~\cite{farsian2020foreground}, reconstructing the lensing potential of CMB maps \cite{caldeira2019deepcmb}, 
estimating full-sky power spectrum from partial-sky spectrum~\citep{Pal:2022hpi} to name a few.

In the context of inpainting in CMB, some of the earlier work includes implementing partial CNN~\cite{montefalcone2021inpainting} to 
inpaint a CMB map masked using a generated mask which covers up to 25\% of the input image. 
In this method, the authors mask the convolutional matrix itself and the inpainting is achieved by using 
only the non-masked pixels.
In another study modified Generative Adversarial Network (GAN)~\cite{Sadr:2020rje} is used to inpaint the masked CMB map. 
This method provides results comparable to the Gaussian constraint realizations method while in a 
non-Gaussian scenario they achieve better results.
Another study by~\cite{yi2020cosmo} used a variational auto-encoder to reconstruct CMB maps which apart 
from inpainting the masked regions can also provide an uncertainty estimate on its predictions.  

In the current work, we aim to develop a network that can inpaint images with large, irregular missing regions, such as a CMB map masked by the Planck 2018 common mask~\citep{Planck:2018yye} 
—which primarily obscures the galactic plane thereby removing about 20\% of the observed sky. 
We introduce a cross-resolution contextual integration framework that seamlessly fuses global features with local fine details in the image.
Dilated convolutions~\citep{yu2015multi}, in our network,  provide a higher receptive field to efficiently capture broader contextual information over larger areas. By combining with the local feature maps from standard convolution operations, our network provides a robust and effective solution for inpainting. 
To further improve the structural integrity of the predictions, we introduce an additional loss function during training that penalizes the structural dissimilarity between the inpainted predicted map and true maps, to the mean squared loss. The combination of these losses enables the network to achieve more accurate and visually coherent reconstructions.

This paper is organized as follows. We describe the proposed network architecture in detail, including a 
brief discussion on dilated convolution layers and our loss function used to train our network in section~\ref{sec:Network Architecture}. In section \ref{sec:dataset}, 
we discuss the dataset generation procedure and provide details about the masks that we use in this work. 
We explain the training procedure in  section~\ref{sec:method}, followed by section~\ref{sec:results}, 
where we present the results from our testing dataset. 
Finally, we conclude with a discussion of the findings and outline  potential directions for 
future work in section \ref{sec:conclusions}.
\begin{figure*}
    \centering
    \includegraphics[width=0.95\textwidth]{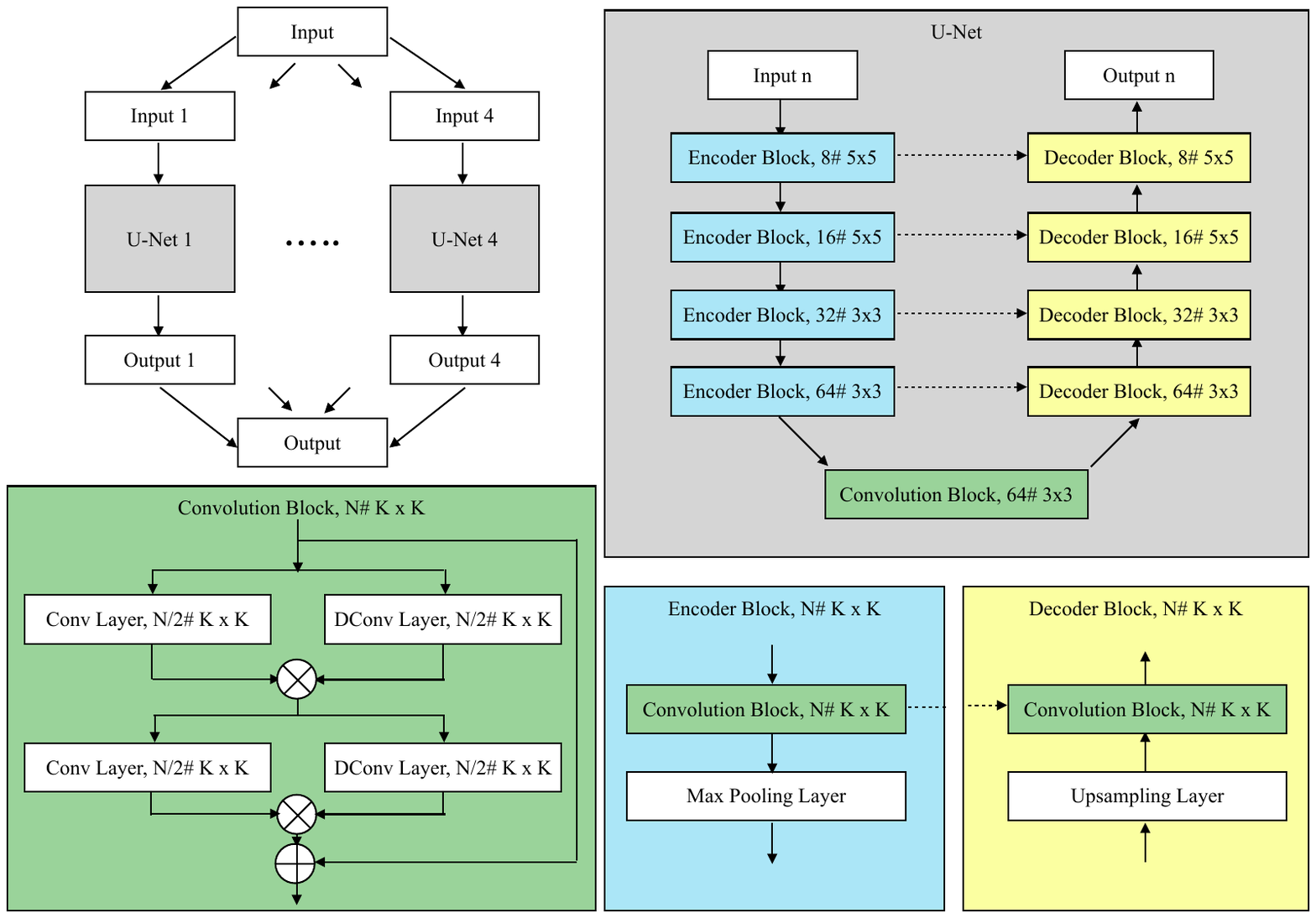}
    \caption{Schematic representaion of SkyReconNet architecture. In top left panel we show the 
     the main network. The compositon of each U-Net sub-network is displayed in the top left panel. There is a 
     separate U-Net for handling images from each region. In the bottom panel from left to right we show the 
     composition of convolutional, encoder, and decoder blocks. }
    \label{fig:blocks_schematic}
\end{figure*}

\section{Network Architecture}\label{sec:Network Architecture}

In this section, we briefly discuss the key features of our network, SkyReconNet, designed specifically 
to inpaint masked regions in a 
CMB map. By leveraging the dilated convolutional layer's expanded receptive field, our network captures global features more effectively. Unlike standard convolutions, dilated convolutions achieve this without increasing the number of parameters, thereby ensuring computational efficiency. 
By merging the feature maps from both dilated and standard convolutional layers, SkyReconNet effectively captures multi-scale information and integrates global features with local details through a cross-resolution approach.

\subsection{Convolutional and Dilated Convolutional Layers}
\label{dl}
For a network with convolutional layers as the feature extracting layers, a convolutional operation is performed 
on the input 2D image as follows,
\begin{equation}
    {\bf G}_{i,j} =\sum^k_{u = -k} \sum^k_{v = -k} H_{u,v} {\bf F}_{i-u,j-v} + b,
\end{equation}
where $b$ is the bias factor, ${\bf F}$ is the input image, $H$ and ${\bf G}$ represents 
the kernel and resulting image after 
convolutional operation respectively. These kernel and bias factors are the learnable parameters that 
will be modified by the network, during the training phase, to minimize the loss function. 

The key advantage of using convolutional layers in a network is the significant reduction in the 
number of network parameters needed 
to process 2D input data as compared to a fully-connected layer. Instead of learning a unique weight for 
each input pixel, the convolutional layers utilize a fixed number of kernels applied uniformly across 
the input 
image. This weight-sharing  mechanism enables the network to leverage the same parameters across different 
image regions, drastically reducing the number of parameters required by the network, 
thereby leading to more efficient computation and better generalization.

However, this advantage comes with a trade-off. While the limited size of the convolutional kernels allows 
the network to learn fine-grained local features more efficiently, it   
constrain the network from capturing global features due to its smaller receptive field. 
This limitation can pose difficulties in identifying overarching structural patterns or large-scale  
features~\citep{hermann2020origins} in the image. To mitigate this, several strategies can be employed. 
One approach is  to implement 
convolutional operations with larger kernel sizes. A larger kernel will have a higher receptive field 
but this comes with computational overhead. Other options such as introducing deeper layers 
or pooling operations are 
generally preferred since they gradually increase the receptive field and capture more abstract, 
high-level information.  
Another approach is to use dilated convolutional layers where the convolutional operation is performed as follows~\citep{yu2015multi},
\begin{equation}
    {\bf D}_{i,j} =\sum^k_{u = -k} \sum^k_{v = -k} H_{u,v} {\bf F}_{i-lu,j-lv} + b \, .
\end{equation}
Here ${\bf D}$ and  $l$ represents the image after a convolutional operation and dilation rate, respectively. 
When we add a dilation with dilation rate $l = n$, $n-1$ pixels will be skipped when calculating the sum, 
resulting in a higher receptive field without increasing the number of parameters. The drawback of using the
dilated convolutional layer is, 
it tends to create blind spots by skipping over some pixels. 

In our work, we use a configuration introduced by~\citep{8502129}, called a mixed layer where we used a 
combination of convolutional and dilated convolutional layers together. While a typical configuration of 
2 convolutional layers can in principle provide a larger receptive field, but doing so will 
lead to narrow receptive field intensity distribution, making it more sensitive to small features. 
Meanwhile, by using a mixed 
layer configuration, one can achieve a larger receptive field without creating blind spots while having 
a broader receptive field intensity distribution, striking a good balance between small and large-scale features. 
In our network, we will use 2 mixed layers followed by a residual connection to form a convolution block, 
as shown in the bottom left panel of figure~\ref{fig:blocks_schematic}. 

\subsection{Network}\label{sec:NArch}
The schematic of our network is shown in figure~\ref{fig:blocks_schematic}. Our network can be characterized 
as a nested framework of $n$ U-Nets~\cite{ronneberger2015u}, where $n$ is the 
total number of regions into which the full-sky 
CMB map is divided into. In the current work, we implemented a strategy introduced by~\citep{Sudevan:2024hwq} for processing the full-sky data. 
First, we divide the full-sky into 4 regions each with a dimension 
$3N_{side} \times N_{side}$, where $N_{side}$ refers to the HEALPix~\cite{Gorski_2005} pixel resolution 
parameter. We construct a separate sub-network to process images corresponding to each of these 4 regions 
as shown in the top left panel of figure~\ref{fig:blocks_schematic}.

Each sub-network consists of a U-Net with an encoder and decoder unit as shown in the top right panel of figure~\ref{fig:blocks_schematic}. The encoder consists of alternating convolution blocks and pooling layers. 
Each convolution block consists of two channels, in channel 1 the input image is processed by convolutional operations with kernel 
size $k \times k$ and the number of filters for each layer is $N/2$. Application of each 
convolutional layer is followed by an activation layer and a Batch Normalization layer. We use 
$parametric$ Rectified Linear Unit ($p$-ReLU) as the activation function. 
In the second channel, we apply a dilated convolution operation with kernel size $k \times k$, dilation rate 2, and 
number of filters $N/2$. Each convolution operation is followed by a $p$-ReLU activation and a 
BatchNormalization layer. The output from both channels is then concatenated and the process is repeated with 
this concatenated output. The resulting convolution block is shown in the bottom left panel of 
figure~\ref{fig:blocks_schematic}.   
As mentioned in section~\ref{dl}, applying a dilated convolutional operation with dilation rate 2 will increase the receptive 
field but comes with the cost of losing local information due to skipping over pixels. But by combining the 
output from channel 1 which in turn is generated by the application of a convolutional layer with dilation rate 1 
negates this loss of 
local information. 
Lastly, a residual connection is applied to the combined output feature maps by adding the 
input image of the convolutional block back to the concatenated output.

The output from the convolutional block is then passed through a Maxpooling layer to reduce the dimension of the image by half and then we repeat the process 
once again. Our encoder unit consists of 4 such convolution blocks and Maxpooling layers. 
The output from each convolution block in the encoder is later used as a skip connection for the corresponding decoder unit. 
In the decoder unit, the input feature maps are first processed by a convolution block followed by an upsampling 
layer. This upsampling increases the feature maps resolution by a factor of 2. The resulting output is 
then combined with the skip connection from corresponding encoder block before it is processed by 
another convolution block. The skip connections improve the network's ability to reconstruct more detailed 
features of the data, which tends to be discarded in the vanilla encoder-decoder structure. This process is 
repeated till the final output has the same resolution as the input.

In the current framework, we have 4 pairs of encoder-decoder units. The output from these 4 U-Nets 
will be combined together for the loss calculation. In practice, the number of U-Nets 
required depends on the mask area and the distribution of masked regions in the sky. In the present work, we 
use the Planck common mask, which masks roughly 20$\%$ of the sky, predominantly galactic regions but also 
some off-galactic regions. Hence in order to faithfully reconstruct the 
full-sky map we use the entire remaining region of the sky. 

\subsection{Loss Function}
The loss function plays a crucial role in the training process of neural networks by  
guiding the network in estimating the optimal weights. We use a composite loss function that 
integrates two distinct losses: the Mean Squared Error (MSE) and the Structural Similarity Index 
Measure (SSIM)~\citep{1284395}. The primary objective of this combined loss function is to minimize not only the 
pixel-to-pixel discrepancies but also to preserve the overall structural features of CMB images.

The Mean Squared Error (MSE) loss function is a widely used metric that quantifies the average squared 
difference between the target values and the predicted values. The MSE loss function is expressed as,
\begin{equation}
    L_{MSE} = \frac{1}{N_i N_j} \sum^{N_i,N_j}_{ij}\left( \textbf{Y}_{ij} -  \hat{\textbf{Y}}_{ij}\right)^2,
\end{equation}
where $\textbf{Y}$ and $\hat{\textbf{Y}}$ denotes the target and  the predicted outputs respectively. 
$N_i$  
and $N_j$ represents the width and height respectively of the image, respectively. The MSE focuses on minimizing the differences 
at each pixel, thereby ensuring the predicted image matches closely to the target image 
in terms of individual pixel values.

In contrast, the Structural Similarity Index Measure (SSIM) is designed to evaluate the perceptual 
quality of images by considering changes in structural information. The SSIM-based loss function~\citep{1284395} is formulated as,
\begin{equation}
    L_{SSIM} = 1 -\frac{(2\mu_\textbf{Y} \mu_{\hat{\textbf{Y}}} + c_1) (2 \sigma_{\textbf{Y}\hat{\textbf{Y}}} +c_2)}{(\mu_\textbf{Y}^2 + \mu_{\hat{\textbf{Y}}}^2 +c_1) (\sigma_\textbf{Y}^2 + \sigma^2_{\hat{\textbf{Y}}} + c_2)},
\end{equation}
where $\mu$ and $\sigma$ are the pixel mean and variance of $\textbf{Y}$ or $\hat{\textbf{Y}}$, 
and $\sigma_{\textbf{Y}\hat{\textbf{Y}}}$ 
is the covariance of $\textbf{Y}$ and $\hat{\textbf{Y}}$. The constants $c1$ and $c2$ are included to stabilize the division 
when the denominators are small. The SSIM loss function focuses on preserving the structural integrity 
of the images, which is important for maintaining the structural properties of CMB data.

The final loss function is a weighted combination of the MSE and SSIM loss functions, allowing us to 
balance the contributions of pixel-wise accuracy and structural similarity. The combined loss function 
is defined as,
\begin{equation}
    L = \alpha L_{MSE} + \beta L_{SSIM}\, ,
\end{equation}
where $\alpha$ and $\beta$ are the weighting coefficients that determine the relative importance of 
the MSE and SSIM components. In our study, we empirically selected $\alpha=5$, $\beta=1$, $c1 = 0.01$ and
$c2$  = 0.03 to achieve 
an optimal balance. By incorporating both MSE and SSIM into the loss function, our approach ensures that 
the network not only minimizes the squared differences between the predicted and target images but 
also preserves the essential structural characteristics of the CMB.

\section{Dataset Generation} \label{sec:dataset}

To train the network, we simulate full-sky CMB maps using publicly available software 
packages CAMB~\citep{lewis2000efficient} and HEALPix~\citep{Gorski_2005}. To generate multiple CMB realizations, 
we sample the cosmological parameters from a normal distribution with mean set as the  
best-fit values provided by Planck 2018 results~\citep{aghanim2020planck} and with 1$\sigma$ standard 
deviation. The cosmological parameters we sample are as follows: $H_0 = 67.37 \pm 0.54~km~s^{-1}~Mpc^{-1}$, 
$\Omega_b h^2 = 0.02233 \pm 0.00015 $, $\Omega_c h^2 = 0.1198 \pm 0.0012$, $\tau = 0.0540 \pm0.0074$,
$10^9$,  $A_s = 2.105 \pm0.030$ and $n_s = 0.9652 \pm 0.0042 $. With each sampled set of cosmological parameters as inputs to CAMB, we generate corresponding
angular power spectrum, which is then used to generate 
a full-sky map with HEALPix at a pixel resolution defined by HEALPix pixel resolution parameter $N_{side}= 64$. Afterwards, 
we apply a Gaussian beam smoothing with a full-width half maximum $0.92^\circ$. In this way, 
we generate a total of 1200 full-sky simulations. 

The CMB simulations generated using HEALPix is a 1D array and since CNN generally processes 
data in the form of 2D images, we need to transform the full-sky maps from HEALPix 1D to 2D. To do 
so, we first reorder all the maps from the HEALPix RING pixelation in which they are created to the NESTED scheme.
We then divide maps in the NESTED scheme into 12 equal areas with $N_{side} \times N_{side}$ dimension. 
Afterward, we use the approach introduced in~\citep{Sudevan:2024hwq}, where we take one area in 
the low-latitude region and combine it with the 2 neighboring higher-latitude 
areas. 
This creates 4 larger areas with $3N_{side} \times N_{side}$ dimension. The motivation for including all the remaining regions of 
the sky is because we use the Planck 2018 common mask to remove 
about 20$\%$ of the total sky, mostly the galactic region and some off-galactic regions.

In order to form the input dataset,  we reformat the 1D full-sky CMB maps to four 
$3N_{side} \times N_{side}$ planar images, afterward we normalize the planar maps by dividing the pixel 
values by the corresponding full-sky map's standard deviation.  We then form the input 
dataset by multiplying the normalized CMB planar maps with the planar mask maps. The output target 
dataset comprises of the unmasked original CMB maps.
\begin{figure}
    \centering
    \includegraphics[width=0.5\textwidth]{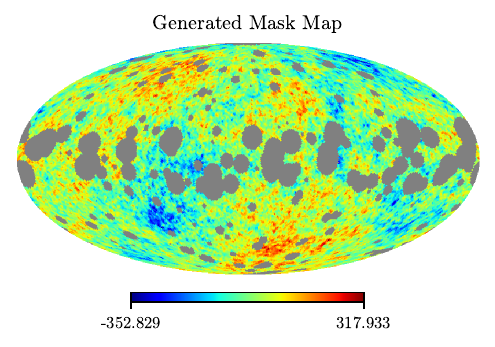}
    \caption{An input CMB map convolved with our custom generated mask created by combining 250 
    circular regions with random positions and sizes. The circular masks in the galactic
    region are assigned larger radii as compared to the outer regions. This mask removes  
    approximately 21$\%$ of the sky, with 
    the gray pixels indicating the masked regions.}
    \label{fig:mask_map1}
\end{figure}

\begin{figure}
    \centering
    \includegraphics[width=0.5\textwidth]{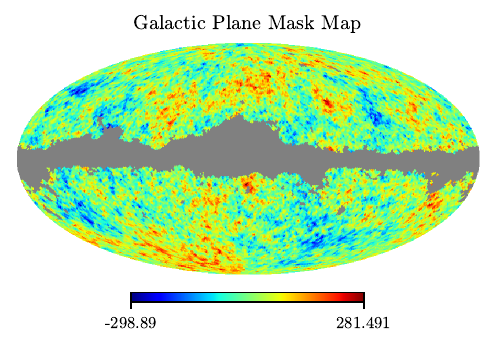}
    \caption{We show an input CMB map multipiled by Planck common mask provided by Planck science team. 
    This mask predominantly removes the galactic region and excludes roughly $20\%$ of the total sky. The grey 
    pixels corresponds to the masked regions.}
    \label{fig:mask_map2}
\end{figure}

During training, out of the total 1200 pairs of masked input CMB and unmasked target CMB maps, pairs of 200 simulations 
are set as the testing dataset, 150 as the validation dataset, and the rest forms the training dataset.

\begin{figure*}
    \centering
    \includegraphics[width=\textwidth]{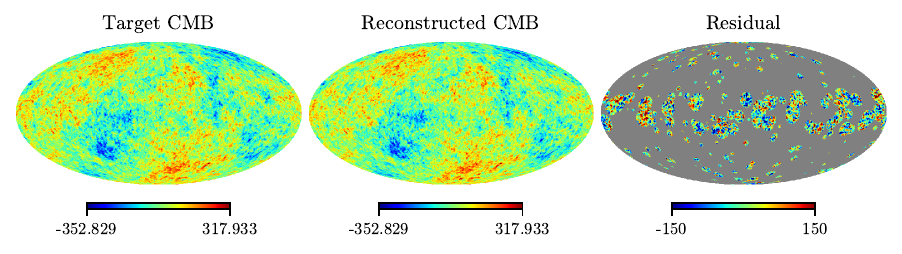}
    \caption{The left panel is the target CMB temperature map, taken from the testing dataset which 
    the network had not seen during training. The reconstructed CMB temperature map generated by our network 
    trained on input maps masked by our custom generated mask is displayed in the middle panel.  
    The reconstructed CMB map closely resembles the target CMB map.  In the right panel, we show the 
    residual temperature map, obtained after subtracting the reconstructed from the target CMB temperature 
    maps. All maps are shown in scale of $\mathrm{\mu K}$.}
    \label{fig:Result1}
\end{figure*}

\begin{figure*}
    \centering
    \includegraphics[width=0.9\textwidth]{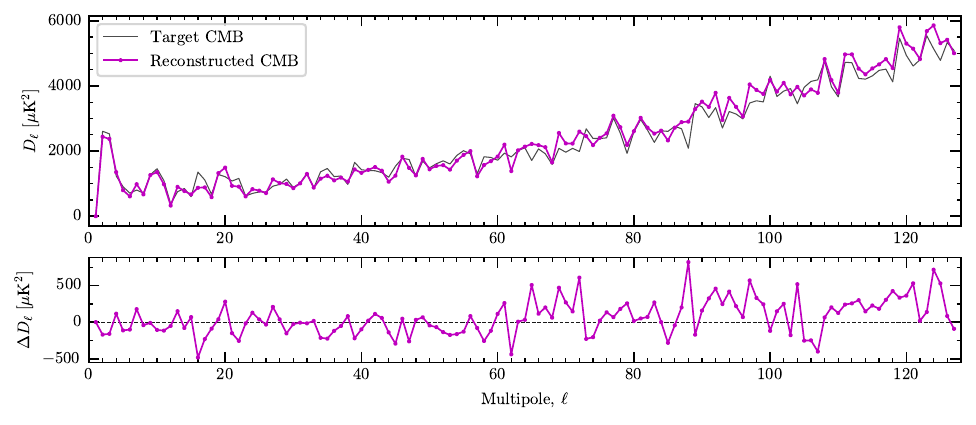}
    \caption{    
    The angular temperature power spectrum of the estimated from our network predicted full-sky 
    CMB map, shown with the magenta line in the top panel. 
    The input CMB map is masked by our custom generated mask. 
    The angular power spectrum of the target CMB map is shown in black. We can see both the power spectra are 
    closely matched. The difference between these two power spectra is shown in the bottom panel.
    }
    \label{fig:Dl1}
\end{figure*}
\begin{figure*}
    \centering
    \includegraphics[width=1.\textwidth]{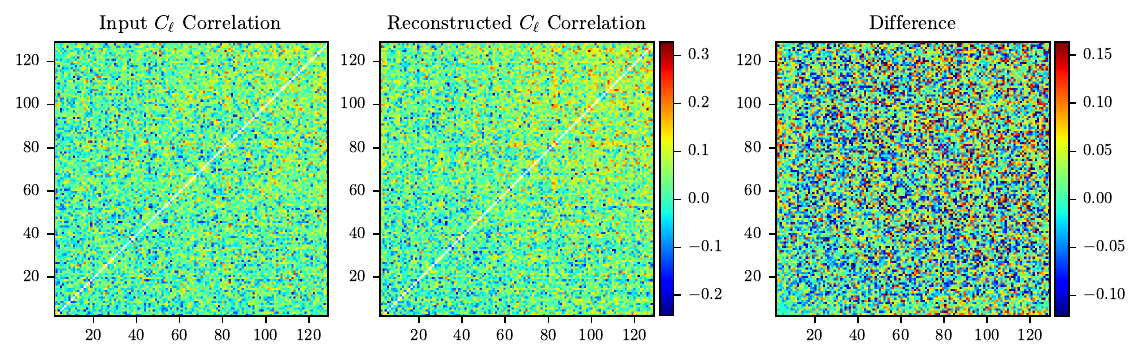}
    \caption{    
    The left panel displays the covariance matrix computed from 150 angular power spectra corresponding to  
    target CMB maps from the testing dataset. This matrix represents the correlations between pairs of 
    multipoles $\ell$ and $\ell^{\prime}$. The covariance matrix estimated from our network predicted full-sky 
    CMB maps angular power spectra is shown in the middle panel. Visually, both covariance matrices match 
    closely. There is no sign of any spurious correlations existing between any pairs of different multipoles 
    ($\ell$, $\ell^{\prime}$). This is further evidenced by the difference plot, obtained after subtracting the 
    covariance matrix corresponding to the predicted CMB maps from that of the target CMB maps.
    }
    \label{fig:cov1}
\end{figure*}

\subsection{Masks}
In this work, we use two different masks, a custom generated mask and a galactic mask  provided by the Planck science team. 
The generated mask is formed by combining 150 randomly positioned circular masks, with a randomized 
radius up to 4 pixels, on the region $30^{\circ}$ 
above and below the 
galactic plane. For the galactic plane (within $\pm 30^{\circ}$), 
we reduce the 
number of circles to 100 while 
increasing their radius up to 10 pixels. In total, we will have 250 circular masks on the map, 
that cover $\sim$21\% of the sky. For the Planck mask, we use the Planck 2018 common mask~\citep{Planck:2018yye}, generated by combining the results of all component separation methods used by the Planck science team 
and eliminating pixels with high standard deviation. 
We show these masks  in figures~\ref{fig:mask_map1} and~\ref{fig:mask_map2} after multiplying the 
respective masks with the target CMB map shown in figure~\ref{fig:Result1}.

\section{Methodology}\label{sec:method}
To evaluate the efficiency of our network in refilling the missing information in a partial-sky CMB map, 
we train the network using two different masks. In the first analysis, the input maps used to train the network are partial-sky CMB maps 
generated by masking the full-sky 
maps with our mask. This procedure removes about $\sim$21$\%$ of the total sky area   
and we compare the predicted inpainted output maps against the true 
full-sky maps. In the second analysis, we use the Planck 2018 common mask to generate input partial-sky maps. During training, the network 
estimates the optimal weights by minimizing a weighted linear combination of two 
loss functions, the MSE and SSIM. The weighting factors in this combined loss function are determined empirically. We use Adam optimization
scheme~\citep{kingma2014adam} initialized with a learning rate set to 0.0001 while training the network. The learning rate is 
gradually reduced by 25$\%$ over the course of training 
if the validation loss is not improved over consecutive 50 
training epochs. The lower bound for the learning rate is set at $10^{-6}$ in our analysis. 
We provide the training data in batches to the network with a batch size of 25 and 10000 training epochs.
We set an early stopping condition that if the validation loss does not improve for about 250 consecutive 
epochs once the learning rate is $10^{-6}$, then the network will stop training. The final optimal weight 
corresponds to the set of weights for which the validation loss is the minimum during 
the training phase. We use the testing dataset 
to evaluate the performance of our network by comparing the predicted inpainted CMB maps with the actual 
full-sky CMB maps for every example in the testing dataset. We also compare the power spectra estimated from
both maps. 
The results of both analyses are presented in Section~\ref{results_sm} and~\ref{results_pl}.

\begin{figure}
    \centering
    \includegraphics[width=0.45\textwidth, height=0.3\textwidth]{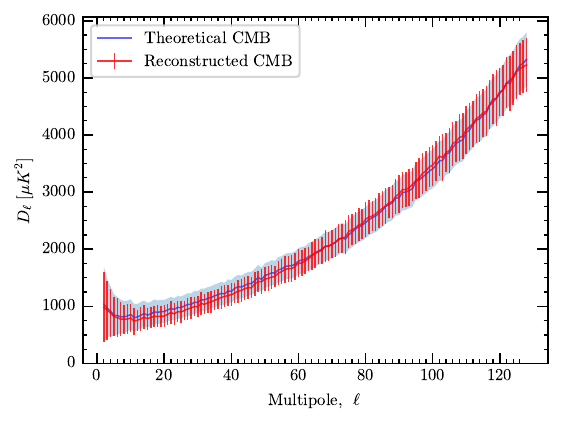}
    \caption{    
    The angular power spectrum averaged over the 150 reconstructed CMB maps and their corresponding 
    target CMB maps from the testing dataset is displayed as red and black lines, respectively. The error bars 
    in red indicate the standard deviation of the angular power spectra predicted by our network across all 
    200 examples at each multipoles,  $\ell$.  These error bars provide a direct measure of the variability 
    in our network's predictions. Note that the error bars closely match the 
    cosmic variance, shown as a blue band.
    }
    \label{fig:comb_Dl1}
\end{figure}

\section{Results}\label{sec:results}
We discuss the results obtained after training our SkyReconNet in this section. As mentioned in 
Section~\ref{sec:method} we consider two cases. In case 1, we train the network to inpaint a partial-sky 
map created by masking the full-sky map with a mask generated by us. While in case 2, we use Planck 2018 
common mask to generate partial-sky maps. 
\subsection{Results with the Generated Mask Map}
\label{results_sm}
\begin{figure*}
    \centering
    \includegraphics[width=\textwidth]{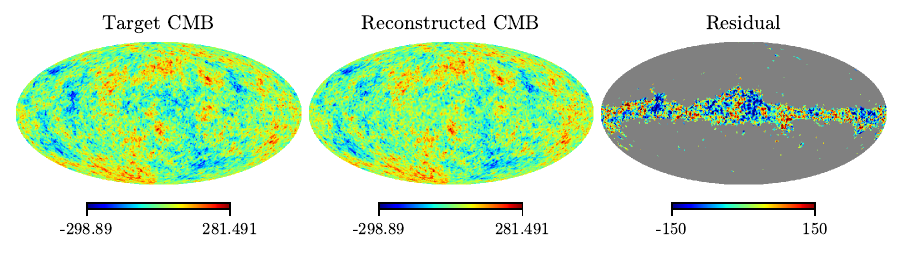}
    \caption{The left panel is a target CMB temperature map from the testing dataset with the input full-sky 
    CMB maps masked by Planck 2018 common mask. The reconstructed CMB temperature map as predicted by our 
    trained network after inpainting the missing information in the regions excluded by the Planck mask is shown in 
    the middle panel. On the right we display the residual map after taking the difference between 
    the reconstructed and target full-sky CMB maps. All maps are shown in scale of $\mathrm{\mu K}$.}
    \label{fig:Result2}
\end{figure*}
\begin{figure*}
    \centering
    \includegraphics[width=0.9\textwidth]{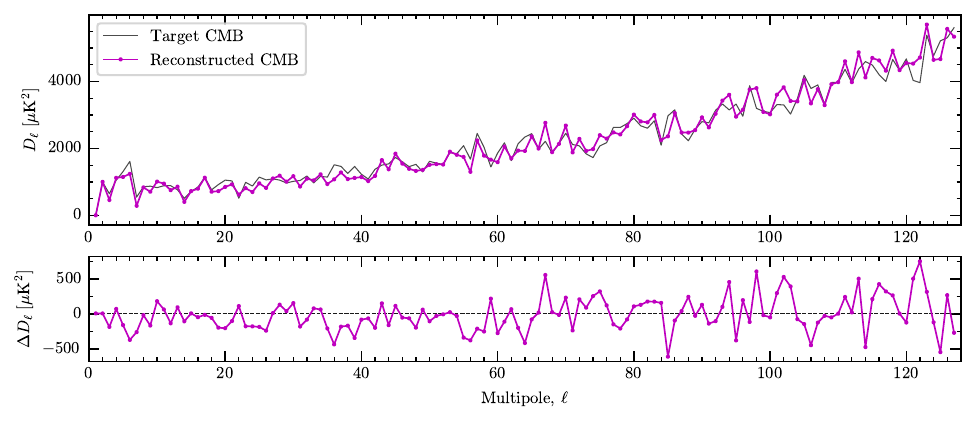}
    \caption{    
    In the top panel, the angular temperature power spectrum estimated from the recovered CMB map and the target 
    CMB map is shown in magenta and black lines respectively. Both power spectrum matches quite well and there is 
    no sign of any positive or negative bias at higher multipoles $\ell$. The difference between the two 
    power spectra is shown in the bottom panel.
    }
    \label{fig:Dl2}
\end{figure*}
\begin{figure*}
    \centering
    \includegraphics[width=1.\textwidth]{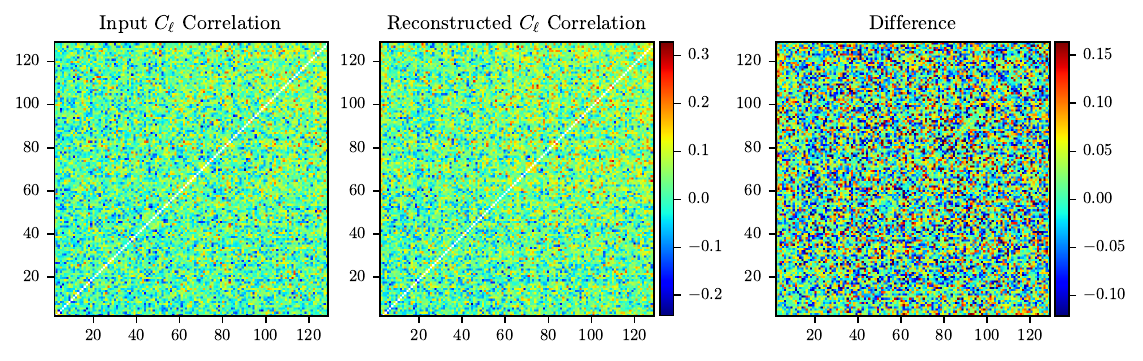}
    \caption{    
    The covariance matrix computed from 150 angular power spectra corresponding to  
    target CMB maps from the testing dataset is shown in the left panel. 
    The covariance matrix estimated from our full-sky 
    CMB maps angular power spectra that our network provided after inpainting the missing regions in 
    the Planck 2018 common mask is shown in the middle panel and the difference between these two matrices is shown in 
    the right panel. We see a close match between these two matrices and the difference between them shows that
    there is no unwanted correlations between any pairs of multipoles ($\ell$, $\ell^{\prime}$).
    }
    \label{fig:cov2}
\end{figure*}

\begin{figure}
    \centering
    \includegraphics[width=0.45\textwidth, height=0.3\textwidth]{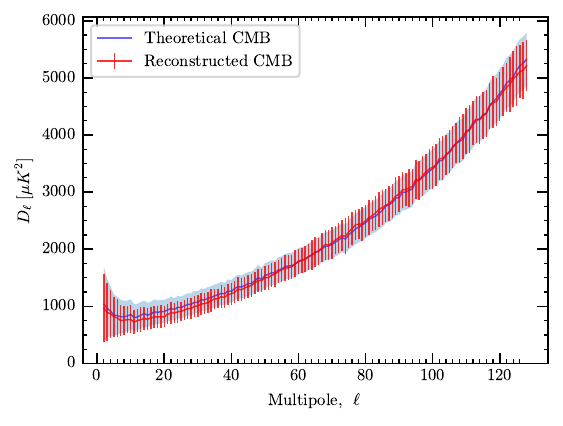}
    \caption{ 
    The mean angular power spectrum estimated from the 150 reconstructed CMB maps and their corresponding 
    target CMB maps from the testing dataset is displayed as red and black lines, respectively. The error bars 
    in red indicate the standard deviation of the angular power spectra predicted by our network. 
    The error bars closely match the cosmic variance, shown as a blue band. 
        }
    \label{fig:comb_Dl2}
\end{figure}

In this section, we present the results after training our network using partial-sky CMB created 
using our custom mask as the input. We use the testing 
dataset, which was not exposed to the network during the training phase, to evaluate the performance of our trained network. 
This ensures an unbiased assessment of its ability to reconstruct the missing regions in the input maps.
In the left panel of the  figure~\ref{fig:Result1} 
displays an example of the target CMB map from the testing dataset. The reconstructed CMB map after inpainting 
the masked regions using our network is shown in the middle panel. The predicted map closely resembles 
the structure of the input map, highlighting the network's efficacy in preserving the critical features in the 
CMB map. 
To qualitatively evaluate the reconstruction accuracy, we compute the residual map by
taking the difference between our predicted map and the 
input CMB map. The residual map is shown in the left panel of figure~\ref{fig:Result1}. The results demonstarte that our network is able to 
effectively refill the missing regions, as evidenced by the minimal residuals in the difference map. 

Using the target and the predicted CMB maps, we compute the full-sky angular power spectrum $C_\ell$ using 
the HEALPix Anafast module, with maximum multipole sat at $\ell_{max} = 2 N_{side}$ after properly taking into 
account of the beam and pixel window 
functions. The resulting power spectra and their difference is presented in the top and bottom panel of figure~\ref{fig:Dl1}, respectively.  The black line represents the 
target CMB power spectrum while the magenta line corresponds to the reconstructed CMB power spectrum predicted 
by our network. We see a close agreement between both the power spectra.  
From the bottom panel of figure~\ref{fig:Dl1}, the difference between the two power spectra shows no 
evidence of the presence of any negative bias at higher multipoles, $\ell$. This indicates that our network is 
able to effectively reconstruct the fine-scale structures of the CMB without introducing any significant 
errors or systematic deviations, even at smaller angular scales.

To further validate our method, we use 150 true CMB maps from the testing dataset to estimate their 
corresponding angular power spectra. These spectra are then utilized to compute   
the covariance matrix, which captures the statistical correlation between different multipole 
pairs ($\ell$, $\ell^{\prime}$). The covariance matrix derived from the true CMB maps is 
shown in the left panel of figure~\ref{fig:cov1}. In the middle panel, we display the covariance matrix 
estimated from the inpainted CMB maps predicted by our 
network using the same testing dataset. The right panel presents the difference between the two 
covariance matrices. We see our network predicted CMB maps preserve the covariance structure across all 
multipole $\ell$ and $\ell^{\prime}$ moments. This demonstrates that the  
inpainting procedure do not introduce any spurious correlations between any multipole moments. 

In figure~\ref{fig:comb_Dl1}, we show the mean of all the angular power spectra estimated from 
all the 150 true CMB maps 
in the testing dataset and reconstructed CMB maps in black and red lines respectively.  
Both power spectra exhibit excellent agreement across all multipoles. The blue band in the 
figure~\ref{fig:comb_Dl1}, represents the  
cosmic variance at each multipoles $\ell$, which sets a fundamental limit on power spectrum uncertainties. 
We estimate the standard deviation using all 150 predicted CMB map power spectra for each multipoles $\ell$ and is 
shown as the light red vertical lines in figure~\ref{fig:comb_Dl1}.  This close match suggests that the 
variability in the reconstructed CMB maps is consistent with the fundamental limits imposed by 
cosmic variance, rather than being dominated by systematic errors or overfitting from the network.
These results demonstrate our network's ability to reliably inpaint CMB maps while preserving the 
statistical properties inherent to the cosmological signal.

\subsection{Results with Planck 2018 Common Mask}
\label{results_pl}

We present the predictions of our network trained on CMB maps masked with the Planck 2018 common mask. This 
mask predominantly excludes  
the central galactic region. To illustrate the performance of our network in inpainting a CMB map masked by 
Planck 2018 common mask, we display an example from the testing dataset in figure~\ref{fig:Result2}. The left panel 
showcases the target CMB map from the testing dataset which is then convolved with the Planck 2018 common mask and 
serves as the input to our trained network. The middle panel displays the predicted full-sky CMB map, 
where our network has inpainted the regions excluded by the Planck mask. From visual inspection, it is 
evident that our network is able to efficiently reconstruct the 
CMB information in the galactic region. Finally, we show the residual map 
after subtracting our reconstructed CMB map from the target CMB map in the right panel of figure~\ref{fig:Result2}.

Similar to the analysis carried out in  Section~\ref{results_sm}, we compute the angular power spectra for 
all the target maps in the testing dataset and corresponding our network predicted CMB maps. 
We display both the target and the corresponding predicted CMB map 
power spectra in the top panel of the figure~\ref{fig:Dl2}. We see that the predicted power spectrum 
closely matches the target, with no evidence of any significant positive or negative bias at any multipole range. 
The bottom panel of figure~\ref{fig:Dl2} quantifies the difference between the two power spectra.

To compare the covariance properties, we computed the covariance matrices corresponding to the target and 
predicted CMB angular power spectra. In the left and middle panels of figure~\ref{fig:cov2}, we show the  
covariance matrices from target and predicted maps, respectively. 
We observe that the covariance matrix obtained from our network predictions exhibits no significant 
correlations between any pair of multipole moments $\ell$ 
and $\ell^{\prime}$. The covariance structure mirrors that of the target CMB covariance matrix. 
This similarity is  evidenced by the difference plot shown in the right panel of figure~\ref{fig:cov2}.  
In figure~\ref{fig:comb_Dl2}, we show the mean angular power spectra from the target and the predicted CMB 
maps in blue and red lines respectively. The two power spectra are in excellent agreement with each other. 
The estimated standard deviation from the predicted CMB maps power spectra 
aligns well with the cosmic variance at all multipoles $\ell$. 

Overall the result from training with the Planck common mask and our generated mask shows the robustness of  
our network in inpainting a masked CMB map, where the mask area is over 20$\%$ the total sky. 
Our custom mask which is a composite mask of several smaller masks distributed 
randomly throughout the sky and the Planck 2018 common mask on the other hand predominantly 
excludes the entire galactic region altogether with some small off-galactic regions. 
Our network is able is handle both masks and is able to reconstruct the 
missing pixel values excluded by both masks while preserving the angular power 
spectrum and the covariance structure. The power spectrum estimated from the predicted CMB maps 
from both analyses do not suffer from any negative or positive bias. The close match between the 
standard deviation errors and the cosmic variance  
indicates that our network's prediction in inpainting the CMB map is limited only by errors induced by 
the unavoidable cosmic variance.

\section{Conclusions}\label{sec:conclusions}

In this work, we introduce SkyReconNet, a neural network based on U-net architecture enhanced with 
dilated convolutional layers, to reconstruct the missing information in a partial-sky CMB map. 
By leveraging the expanded receptive field of  
dilated convolutional layers with the dilation rate set to 2 along with the detailed local information captured 
by the standard convolutional layers, our network effectively and 
accurately inpaints the masked regions in the sky. 
The capability of our network is further boosted by incorporating  a composite loss function, which is a 
weighted linear 
combination of Structural Similarity Index Measure (SSIM) to  
preserve the structural characteristics, and Mean Squared Error (MSE) to minimize the pixel-level differences. 
Together, by utilizing the expanded receptive field of dilated convolutional layers and this loss 
function, enables our network to maintain the integrity of the reconstructed full-sky map.

To train the network we have simulated 1200 simulations of CMB using publicly available 
software packages CAMB and HEALPix. Each simulated CMB map corresponds to the cosmological parameters 
sampled from Planck 2018 best-fit values.   
Two distinct masks were employed to evaluate our network's efficiency in handling different types of masks. We 
use Planck 2018 common mask, which primarily obscures the central galactic region 
and covers $\sim$20$\%$ of the sky, and a custom-generated mask comprising 250 randomly placed circular 
regions that collectively mask $\sim$21$\%$ of the sky.

Our network is initialized using  Adam optimization
scheme with the initial learning rate set as 0.0001 with a lower bound $10^{-6}$. The learning rate is 
gradually reduced by 25$\%$ if the validation loss fails to improve over 50 consecutive training epochs. 
During training our network 
minimizes the combined loss function, to estimate 
the optimal weights. 

Once trained, we use the testing dataset to demonstrate the efficiency of our network in inpainting 
masked regions in the Planck 2018 common mask and our custom generated mask. Map level and power spectrum  
comparison reveal that our network's predictions closely align with the original full-sky maps.  
The power spectrum estimated from the inpainted maps do not show any signs of positive or negative bias. 
Furthermore, the 
covariance matrix analysis confirms that our network predictions do not lead to any spurious correlations 
between any different pairs of multipole moments.  
Overall, the results establish that our SkyReconNet as a powerful tool for inpainting CMB maps, 
particularly when preserving the structural integrity of CMB is critical. 

Our network focuses on learning statistical features in the CMB map to reconstruct the missing pixels. It aims
to avoid introducing any additional sources of bias or variance beyond those inherently imposed by cosmic
variance. In doing so, it maintains statistical consistency with the underlying CMB statistics, providing a
valuable tool in CMB analyses.

In summary, our network demonstrates exceptional performance in reconstructing the missing information in 
partial-sky CMB maps while ensuring that the inpainted maps remain consistent with the true cosmological 
signal, both in terms of power spectrum and covariance structure. This makes our network a valuable tool 
for inpainting missing data in CMB analyses. 
It would be interesting to see possible extensions of this 
approach to the CMB polarization maps and to experiments in other fields that generates image-like data 
often plagued by missing or corrupted pixels. Our method presents a promising and exciting avenue for future explorations.

\begin{acknowledgments}
This work is based on observations obtained with Planck
(http://www.esa.int/Planck). Planck is an ESA science mission with instruments 
and contributions directly funded by
ESA Member States, NASA, and Canada. We acknowledge the use of Planck Legacy Archive (PLA). 
We use the publicly available HEALPix~\citep{Gorski_2005} package 
(http://healpix.sourceforge.net) for the
analysis of this work. The network we have developed is
based on the libraries provided by Tensorflow, although the same
can be designed using other ML-based platforms.
\end{acknowledgments}


\bibliography{main}

\begin{thebibliography}{85}%
\makeatletter
\providecommand \@ifxundefined [1]{%
 \@ifx{#1\undefined}
}%
\providecommand \@ifnum [1]{%
 \ifnum #1\expandafter \@firstoftwo
 \else \expandafter \@secondoftwo
 \fi
}%
\providecommand \@ifx [1]{%
 \ifx #1\expandafter \@firstoftwo
 \else \expandafter \@secondoftwo
 \fi
}%
\providecommand \natexlab [1]{#1}%
\providecommand \enquote  [1]{``#1''}%
\providecommand \bibnamefont  [1]{#1}%
\providecommand \bibfnamefont [1]{#1}%
\providecommand \citenamefont [1]{#1}%
\providecommand \href@noop [0]{\@secondoftwo}%
\providecommand \href [0]{\begingroup \@sanitize@url \@href}%
\providecommand \@href[1]{\@@startlink{#1}\@@href}%
\providecommand \@@href[1]{\endgroup#1\@@endlink}%
\providecommand \@sanitize@url [0]{\catcode `\\12\catcode `\$12\catcode
  `\&12\catcode `\#12\catcode `\^12\catcode `\_12\catcode `\%12\relax}%
\providecommand \@@startlink[1]{}%
\providecommand \@@endlink[0]{}%
\providecommand \url  [0]{\begingroup\@sanitize@url \@url }%
\providecommand \@url [1]{\endgroup\@href {#1}{\urlprefix }}%
\providecommand \urlprefix  [0]{URL }%
\providecommand \Eprint [0]{\href }%
\providecommand \doibase [0]{http://dx.doi.org/}%
\providecommand \selectlanguage [0]{\@gobble}%
\providecommand \bibinfo  [0]{\@secondoftwo}%
\providecommand \bibfield  [0]{\@secondoftwo}%
\providecommand \translation [1]{[#1]}%
\providecommand \BibitemOpen [0]{}%
\providecommand \bibitemStop [0]{}%
\providecommand \bibitemNoStop [0]{.\EOS\space}%
\providecommand \EOS [0]{\spacefactor3000\relax}%
\providecommand \BibitemShut  [1]{\csname bibitem#1\endcsname}%
\let\auto@bib@innerbib\@empty
\bibitem [{\citenamefont {Bennett}\ \emph
  {et~al.}(2003{\natexlab{a}})\citenamefont {Bennett} \emph
  {et~al.}}]{WMAP:2003ivt}%
  \BibitemOpen
  \bibfield  {author} {\bibinfo {author} {\bibfnamefont {C.~L.}\ \bibnamefont
  {Bennett}} \emph {et~al.} (\bibinfo {collaboration} {WMAP}),\ }\href
  {\doibase 10.1086/377253} {\bibfield  {journal} {\bibinfo  {journal}
  {Astrophys. J. Suppl.}\ }\textbf {\bibinfo {volume} {148}},\ \bibinfo {pages}
  {1} (\bibinfo {year} {2003}{\natexlab{a}})},\ \Eprint
  {http://arxiv.org/abs/astro-ph/0302207} {arXiv:astro-ph/0302207} \BibitemShut
  {NoStop}%
\bibitem [{\citenamefont {Ade}\ \emph {et~al.}(2011)\citenamefont {Ade} \emph
  {et~al.}}]{2011A&A...536A...1P}%
  \BibitemOpen
  \bibfield  {author} {\bibinfo {author} {\bibfnamefont {P.~A.~R.}\
  \bibnamefont {Ade}} \emph {et~al.},\ }\href {\doibase
  10.1051/0004-6361/201116464} {\bibfield  {journal} {\bibinfo  {journal}
  {Astronomy \& Astrophysics}\ }\textbf {\bibinfo {volume} {536}},\ \bibinfo
  {eid} {A1} (\bibinfo {year} {2011})},\ \Eprint
  {http://arxiv.org/abs/1101.2022} {arXiv:1101.2022 [astro-ph.IM]} \BibitemShut
  {NoStop}%
\bibitem [{\citenamefont {Durrer}(2015)}]{Durrer:2015lza}%
  \BibitemOpen
  \bibfield  {author} {\bibinfo {author} {\bibfnamefont {R.}~\bibnamefont
  {Durrer}},\ }\href {\doibase 10.1088/0264-9381/32/12/124007} {\bibfield
  {journal} {\bibinfo  {journal} {Class. Quant. Grav.}\ }\textbf {\bibinfo
  {volume} {32}},\ \bibinfo {pages} {124007} (\bibinfo {year} {2015})},\
  \Eprint {http://arxiv.org/abs/1506.01907} {arXiv:1506.01907 [astro-ph.CO]}
  \BibitemShut {NoStop}%
\bibitem [{\citenamefont {Abazajian}\ \emph {et~al.}(2015)\citenamefont
  {Abazajian} \emph {et~al.}}]{Abazajian:2013vfg}%
  \BibitemOpen
  \bibfield  {author} {\bibinfo {author} {\bibfnamefont {K.~N.}\ \bibnamefont
  {Abazajian}} \emph {et~al.},\ }\href {\doibase
  10.1016/j.astropartphys.2014.05.013} {\bibfield  {journal} {\bibinfo
  {journal} {Astropart. Phys.}\ }\textbf {\bibinfo {volume} {63}},\ \bibinfo
  {pages} {55} (\bibinfo {year} {2015})},\ \Eprint
  {http://arxiv.org/abs/1309.5381} {arXiv:1309.5381 [astro-ph.CO]} \BibitemShut
  {NoStop}%
\bibitem [{\citenamefont {Ade}\ \emph {et~al.}(2014{\natexlab{a}})\citenamefont
  {Ade} \emph {et~al.}}]{Planck:2013jfk}%
  \BibitemOpen
  \bibfield  {author} {\bibinfo {author} {\bibfnamefont {P.~A.~R.}\
  \bibnamefont {Ade}} \emph {et~al.} (\bibinfo {collaboration} {Planck}),\
  }\href {\doibase 10.1051/0004-6361/201321569} {\bibfield  {journal} {\bibinfo
   {journal} {Astron. Astrophys.}\ }\textbf {\bibinfo {volume} {571}},\
  \bibinfo {pages} {A22} (\bibinfo {year} {2014}{\natexlab{a}})},\ \Eprint
  {http://arxiv.org/abs/1303.5082} {arXiv:1303.5082 [astro-ph.CO]} \BibitemShut
  {NoStop}%
\bibitem [{\citenamefont {Aghanim}\ \emph {et~al.}(2020)\citenamefont
  {Aghanim}, \citenamefont {Akrami}, \citenamefont {Ashdown}, \citenamefont
  {Aumont}, \citenamefont {Baccigalupi}, \citenamefont {Ballardini},
  \citenamefont {Banday}, \citenamefont {Barreiro}, \citenamefont {Bartolo},
  \citenamefont {Basak} \emph {et~al.}}]{aghanim2020planck}%
  \BibitemOpen
  \bibfield  {author} {\bibinfo {author} {\bibfnamefont {N.}~\bibnamefont
  {Aghanim}}, \bibinfo {author} {\bibfnamefont {Y.}~\bibnamefont {Akrami}},
  \bibinfo {author} {\bibfnamefont {M.}~\bibnamefont {Ashdown}}, \bibinfo
  {author} {\bibfnamefont {J.}~\bibnamefont {Aumont}}, \bibinfo {author}
  {\bibfnamefont {C.}~\bibnamefont {Baccigalupi}}, \bibinfo {author}
  {\bibfnamefont {M.}~\bibnamefont {Ballardini}}, \bibinfo {author}
  {\bibfnamefont {A.~J.}\ \bibnamefont {Banday}}, \bibinfo {author}
  {\bibfnamefont {R.}~\bibnamefont {Barreiro}}, \bibinfo {author}
  {\bibfnamefont {N.}~\bibnamefont {Bartolo}}, \bibinfo {author} {\bibfnamefont
  {S.}~\bibnamefont {Basak}},  \emph {et~al.},\ }\href@noop {} {\bibfield
  {journal} {\bibinfo  {journal} {Astronomy \& Astrophysics}\ }\textbf
  {\bibinfo {volume} {641}},\ \bibinfo {pages} {A6} (\bibinfo {year}
  {2020})}\BibitemShut {NoStop}%
\bibitem [{\citenamefont {Masi}\ \emph {et~al.}(2002)\citenamefont {Masi},
  \citenamefont {De~Bernardis}, \citenamefont {De~Troia}, \citenamefont
  {Giacometti}, \citenamefont {Iacoangeli}, \citenamefont {Piacentini},
  \citenamefont {Polenta}, \citenamefont {Ade}, \citenamefont {Mauskopf},
  \citenamefont {Bock} \emph {et~al.}}]{masi2002boomerang}%
  \BibitemOpen
  \bibfield  {author} {\bibinfo {author} {\bibfnamefont {S.}~\bibnamefont
  {Masi}}, \bibinfo {author} {\bibfnamefont {P.}~\bibnamefont {De~Bernardis}},
  \bibinfo {author} {\bibfnamefont {G.}~\bibnamefont {De~Troia}}, \bibinfo
  {author} {\bibfnamefont {M.}~\bibnamefont {Giacometti}}, \bibinfo {author}
  {\bibfnamefont {A.}~\bibnamefont {Iacoangeli}}, \bibinfo {author}
  {\bibfnamefont {F.}~\bibnamefont {Piacentini}}, \bibinfo {author}
  {\bibfnamefont {G.}~\bibnamefont {Polenta}}, \bibinfo {author} {\bibfnamefont
  {P.~A.}\ \bibnamefont {Ade}}, \bibinfo {author} {\bibfnamefont {P.~D.}\
  \bibnamefont {Mauskopf}}, \bibinfo {author} {\bibfnamefont {J.}~\bibnamefont
  {Bock}},  \emph {et~al.},\ }\href@noop {} {\bibfield  {journal} {\bibinfo
  {journal} {Progress in Particle and Nuclear Physics}\ }\textbf {\bibinfo
  {volume} {48}},\ \bibinfo {pages} {243} (\bibinfo {year} {2002})}\BibitemShut
  {NoStop}%
\bibitem [{\citenamefont {Bennett}\ \emph
  {et~al.}(2003{\natexlab{b}})\citenamefont {Bennett}, \citenamefont {Bay},
  \citenamefont {Halpern}, \citenamefont {Hinshaw}, \citenamefont {Jackson},
  \citenamefont {Jarosik}, \citenamefont {Kogut}, \citenamefont {Limon},
  \citenamefont {Meyer}, \citenamefont {Page} \emph
  {et~al.}}]{bennett2003microwave}%
  \BibitemOpen
  \bibfield  {author} {\bibinfo {author} {\bibfnamefont {C.~L.}\ \bibnamefont
  {Bennett}}, \bibinfo {author} {\bibfnamefont {M.}~\bibnamefont {Bay}},
  \bibinfo {author} {\bibfnamefont {M.}~\bibnamefont {Halpern}}, \bibinfo
  {author} {\bibfnamefont {G.}~\bibnamefont {Hinshaw}}, \bibinfo {author}
  {\bibfnamefont {C.}~\bibnamefont {Jackson}}, \bibinfo {author} {\bibfnamefont
  {N.}~\bibnamefont {Jarosik}}, \bibinfo {author} {\bibfnamefont
  {A.}~\bibnamefont {Kogut}}, \bibinfo {author} {\bibfnamefont
  {M.}~\bibnamefont {Limon}}, \bibinfo {author} {\bibfnamefont
  {S.}~\bibnamefont {Meyer}}, \bibinfo {author} {\bibfnamefont
  {L.}~\bibnamefont {Page}},  \emph {et~al.},\ }\href@noop {} {\bibfield
  {journal} {\bibinfo  {journal} {The Astrophysical Journal}\ }\textbf
  {\bibinfo {volume} {583}},\ \bibinfo {pages} {1} (\bibinfo {year}
  {2003}{\natexlab{b}})}\BibitemShut {NoStop}%
\bibitem [{\citenamefont {Hincks}\ \emph {et~al.}(2010)\citenamefont {Hincks},
  \citenamefont {Acquaviva}, \citenamefont {Ade}, \citenamefont {Aguirre},
  \citenamefont {Amiri}, \citenamefont {Appel}, \citenamefont {Barrientos},
  \citenamefont {Battistelli}, \citenamefont {Bond}, \citenamefont {Brown}
  \emph {et~al.}}]{hincks2010atacama}%
  \BibitemOpen
  \bibfield  {author} {\bibinfo {author} {\bibfnamefont {A.}~\bibnamefont
  {Hincks}}, \bibinfo {author} {\bibfnamefont {V.}~\bibnamefont {Acquaviva}},
  \bibinfo {author} {\bibfnamefont {P.~A.}\ \bibnamefont {Ade}}, \bibinfo
  {author} {\bibfnamefont {P.}~\bibnamefont {Aguirre}}, \bibinfo {author}
  {\bibfnamefont {M.}~\bibnamefont {Amiri}}, \bibinfo {author} {\bibfnamefont
  {J.}~\bibnamefont {Appel}}, \bibinfo {author} {\bibfnamefont
  {L.}~\bibnamefont {Barrientos}}, \bibinfo {author} {\bibfnamefont {E.~S.}\
  \bibnamefont {Battistelli}}, \bibinfo {author} {\bibfnamefont
  {J.}~\bibnamefont {Bond}}, \bibinfo {author} {\bibfnamefont {B.}~\bibnamefont
  {Brown}},  \emph {et~al.},\ }\href@noop {} {\bibfield  {journal} {\bibinfo
  {journal} {The Astrophysical Journal Supplement Series}\ }\textbf {\bibinfo
  {volume} {191}},\ \bibinfo {pages} {423} (\bibinfo {year}
  {2010})}\BibitemShut {NoStop}%
\bibitem [{\citenamefont {Gandilo}\ \emph {et~al.}(2016)\citenamefont
  {Gandilo}, \citenamefont {Ade}, \citenamefont {Benford}, \citenamefont
  {Bennett}, \citenamefont {Chuss}, \citenamefont {Dotson}, \citenamefont
  {Eimer}, \citenamefont {Fixsen}, \citenamefont {Halpern}, \citenamefont
  {Hilton} \emph {et~al.}}]{gandilo2016primordial}%
  \BibitemOpen
  \bibfield  {author} {\bibinfo {author} {\bibfnamefont {N.~N.}\ \bibnamefont
  {Gandilo}}, \bibinfo {author} {\bibfnamefont {P.~A.}\ \bibnamefont {Ade}},
  \bibinfo {author} {\bibfnamefont {D.}~\bibnamefont {Benford}}, \bibinfo
  {author} {\bibfnamefont {C.~L.}\ \bibnamefont {Bennett}}, \bibinfo {author}
  {\bibfnamefont {D.~T.}\ \bibnamefont {Chuss}}, \bibinfo {author}
  {\bibfnamefont {J.~L.}\ \bibnamefont {Dotson}}, \bibinfo {author}
  {\bibfnamefont {J.~R.}\ \bibnamefont {Eimer}}, \bibinfo {author}
  {\bibfnamefont {D.~J.}\ \bibnamefont {Fixsen}}, \bibinfo {author}
  {\bibfnamefont {M.}~\bibnamefont {Halpern}}, \bibinfo {author} {\bibfnamefont
  {G.}~\bibnamefont {Hilton}},  \emph {et~al.},\ }in\ \href@noop {} {\emph
  {\bibinfo {booktitle} {Millimeter, Submillimeter, and Far-Infrared Detectors
  and Instrumentation for Astronomy VIII}}},\ Vol.\ \bibinfo {volume} {9914}\
  (\bibinfo {organization} {SPIE},\ \bibinfo {year} {2016})\ pp.\ \bibinfo
  {pages} {372--379}\BibitemShut {NoStop}%
\bibitem [{\citenamefont {Gualtieri}\ \emph {et~al.}(2018)\citenamefont
  {Gualtieri}, \citenamefont {Filippini}, \citenamefont {Ade}, \citenamefont
  {Amiri}, \citenamefont {Benton}, \citenamefont {Bergman}, \citenamefont
  {Bihary}, \citenamefont {Bock}, \citenamefont {Bond}, \citenamefont {Bryan}
  \emph {et~al.}}]{gualtieri2018spider}%
  \BibitemOpen
  \bibfield  {author} {\bibinfo {author} {\bibfnamefont {R.}~\bibnamefont
  {Gualtieri}}, \bibinfo {author} {\bibfnamefont {J.}~\bibnamefont
  {Filippini}}, \bibinfo {author} {\bibfnamefont {P.}~\bibnamefont {Ade}},
  \bibinfo {author} {\bibfnamefont {M.}~\bibnamefont {Amiri}}, \bibinfo
  {author} {\bibfnamefont {S.}~\bibnamefont {Benton}}, \bibinfo {author}
  {\bibfnamefont {A.}~\bibnamefont {Bergman}}, \bibinfo {author} {\bibfnamefont
  {R.}~\bibnamefont {Bihary}}, \bibinfo {author} {\bibfnamefont
  {J.}~\bibnamefont {Bock}}, \bibinfo {author} {\bibfnamefont {J.}~\bibnamefont
  {Bond}}, \bibinfo {author} {\bibfnamefont {S.}~\bibnamefont {Bryan}},  \emph
  {et~al.},\ }\href@noop {} {\bibfield  {journal} {\bibinfo  {journal} {Journal
  of Low Temperature Physics}\ }\textbf {\bibinfo {volume} {193}},\ \bibinfo
  {pages} {1112} (\bibinfo {year} {2018})}\BibitemShut {NoStop}%
\bibitem [{\citenamefont {Li}\ \emph {et~al.}(2019)\citenamefont {Li},
  \citenamefont {Li}, \citenamefont {Liu}, \citenamefont {Li}, \citenamefont
  {Cai}, \citenamefont {Li}, \citenamefont {Zhao}, \citenamefont {Liu},
  \citenamefont {Li}, \citenamefont {Xu} \emph {et~al.}}]{li2019probing}%
  \BibitemOpen
  \bibfield  {author} {\bibinfo {author} {\bibfnamefont {H.}~\bibnamefont
  {Li}}, \bibinfo {author} {\bibfnamefont {S.-Y.}\ \bibnamefont {Li}}, \bibinfo
  {author} {\bibfnamefont {Y.}~\bibnamefont {Liu}}, \bibinfo {author}
  {\bibfnamefont {Y.-P.}\ \bibnamefont {Li}}, \bibinfo {author} {\bibfnamefont
  {Y.}~\bibnamefont {Cai}}, \bibinfo {author} {\bibfnamefont {M.}~\bibnamefont
  {Li}}, \bibinfo {author} {\bibfnamefont {G.-B.}\ \bibnamefont {Zhao}},
  \bibinfo {author} {\bibfnamefont {C.-Z.}\ \bibnamefont {Liu}}, \bibinfo
  {author} {\bibfnamefont {Z.-W.}\ \bibnamefont {Li}}, \bibinfo {author}
  {\bibfnamefont {H.}~\bibnamefont {Xu}},  \emph {et~al.},\ }\href@noop {}
  {\bibfield  {journal} {\bibinfo  {journal} {National Science Review}\
  }\textbf {\bibinfo {volume} {6}},\ \bibinfo {pages} {145} (\bibinfo {year}
  {2019})}\BibitemShut {NoStop}%
\bibitem [{\citenamefont {Hui}\ \emph {et~al.}(2018)\citenamefont {Hui},
  \citenamefont {Ade}, \citenamefont {Ahmed}, \citenamefont {Aikin},
  \citenamefont {Alexander}, \citenamefont {Barkats}, \citenamefont {Benton},
  \citenamefont {Bischoff}, \citenamefont {Bock}, \citenamefont {Bowens-Rubin}
  \emph {et~al.}}]{hui2018bicep}%
  \BibitemOpen
  \bibfield  {author} {\bibinfo {author} {\bibfnamefont {H.}~\bibnamefont
  {Hui}}, \bibinfo {author} {\bibfnamefont {P.}~\bibnamefont {Ade}}, \bibinfo
  {author} {\bibfnamefont {Z.}~\bibnamefont {Ahmed}}, \bibinfo {author}
  {\bibfnamefont {R.}~\bibnamefont {Aikin}}, \bibinfo {author} {\bibfnamefont
  {K.~D.}\ \bibnamefont {Alexander}}, \bibinfo {author} {\bibfnamefont
  {D.}~\bibnamefont {Barkats}}, \bibinfo {author} {\bibfnamefont {S.~J.}\
  \bibnamefont {Benton}}, \bibinfo {author} {\bibfnamefont {C.~A.}\
  \bibnamefont {Bischoff}}, \bibinfo {author} {\bibfnamefont {J.~J.}\
  \bibnamefont {Bock}}, \bibinfo {author} {\bibfnamefont {R.}~\bibnamefont
  {Bowens-Rubin}},  \emph {et~al.},\ }in\ \href@noop {} {\emph {\bibinfo
  {booktitle} {Millimeter, Submillimeter, and Far-Infrared Detectors and
  Instrumentation for Astronomy IX}}},\ Vol.\ \bibinfo {volume} {10708}\
  (\bibinfo {organization} {SPIE},\ \bibinfo {year} {2018})\ pp.\ \bibinfo
  {pages} {75--89}\BibitemShut {NoStop}%
\bibitem [{\citenamefont {Abazajian}\ \emph {et~al.}(2022)\citenamefont
  {Abazajian}, \citenamefont {Abdulghafour}, \citenamefont {Addison},
  \citenamefont {Adshead}, \citenamefont {Ahmed}, \citenamefont {Ajello},
  \citenamefont {Akerib}, \citenamefont {Allen}, \citenamefont {Alonso},
  \citenamefont {Alvarez} \emph {et~al.}}]{abazajian2022snowmass}%
  \BibitemOpen
  \bibfield  {author} {\bibinfo {author} {\bibfnamefont {K.}~\bibnamefont
  {Abazajian}}, \bibinfo {author} {\bibfnamefont {A.}~\bibnamefont
  {Abdulghafour}}, \bibinfo {author} {\bibfnamefont {G.~E.}\ \bibnamefont
  {Addison}}, \bibinfo {author} {\bibfnamefont {P.}~\bibnamefont {Adshead}},
  \bibinfo {author} {\bibfnamefont {Z.}~\bibnamefont {Ahmed}}, \bibinfo
  {author} {\bibfnamefont {M.}~\bibnamefont {Ajello}}, \bibinfo {author}
  {\bibfnamefont {D.}~\bibnamefont {Akerib}}, \bibinfo {author} {\bibfnamefont
  {S.~W.}\ \bibnamefont {Allen}}, \bibinfo {author} {\bibfnamefont
  {D.}~\bibnamefont {Alonso}}, \bibinfo {author} {\bibfnamefont
  {M.}~\bibnamefont {Alvarez}},  \emph {et~al.},\ }\href@noop {} {\bibfield
  {journal} {\bibinfo  {journal} {arXiv preprint arXiv:2203.08024}\ } (\bibinfo
  {year} {2022})}\BibitemShut {NoStop}%
\bibitem [{\citenamefont {Adak}\ \emph {et~al.}(2022)\citenamefont {Adak},
  \citenamefont {Sen}, \citenamefont {Basak}, \citenamefont {Delabrouille},
  \citenamefont {Ghosh}, \citenamefont {Rotti}, \citenamefont
  {Mart{\'\i}nez-Solaeche},\ and\ \citenamefont {Souradeep}}]{adak2022b}%
  \BibitemOpen
  \bibfield  {author} {\bibinfo {author} {\bibfnamefont {D.}~\bibnamefont
  {Adak}}, \bibinfo {author} {\bibfnamefont {A.}~\bibnamefont {Sen}}, \bibinfo
  {author} {\bibfnamefont {S.}~\bibnamefont {Basak}}, \bibinfo {author}
  {\bibfnamefont {J.}~\bibnamefont {Delabrouille}}, \bibinfo {author}
  {\bibfnamefont {T.}~\bibnamefont {Ghosh}}, \bibinfo {author} {\bibfnamefont
  {A.}~\bibnamefont {Rotti}}, \bibinfo {author} {\bibfnamefont
  {G.}~\bibnamefont {Mart{\'\i}nez-Solaeche}}, \ and\ \bibinfo {author}
  {\bibfnamefont {T.}~\bibnamefont {Souradeep}},\ }\href@noop {} {\bibfield
  {journal} {\bibinfo  {journal} {Monthly Notices of the Royal Astronomical
  Society}\ }\textbf {\bibinfo {volume} {514}},\ \bibinfo {pages} {3002}
  (\bibinfo {year} {2022})}\BibitemShut {NoStop}%
\bibitem [{\citenamefont {Collaboration}\ \emph {et~al.}(2023)\citenamefont
  {Collaboration}, \citenamefont {Allys}, \citenamefont {Arnold}, \citenamefont
  {Aumont}, \citenamefont {Aurlien}, \citenamefont {Azzoni}, \citenamefont
  {Baccigalupi}, \citenamefont {Banday}, \citenamefont {Banerji}, \citenamefont
  {Barreiro} \emph {et~al.}}]{litebird2023probing}%
  \BibitemOpen
  \bibfield  {author} {\bibinfo {author} {\bibfnamefont {L.}~\bibnamefont
  {Collaboration}}, \bibinfo {author} {\bibfnamefont {E.}~\bibnamefont
  {Allys}}, \bibinfo {author} {\bibfnamefont {K.}~\bibnamefont {Arnold}},
  \bibinfo {author} {\bibfnamefont {J.}~\bibnamefont {Aumont}}, \bibinfo
  {author} {\bibfnamefont {R.}~\bibnamefont {Aurlien}}, \bibinfo {author}
  {\bibfnamefont {S.}~\bibnamefont {Azzoni}}, \bibinfo {author} {\bibfnamefont
  {C.}~\bibnamefont {Baccigalupi}}, \bibinfo {author} {\bibfnamefont
  {A.}~\bibnamefont {Banday}}, \bibinfo {author} {\bibfnamefont
  {R.}~\bibnamefont {Banerji}}, \bibinfo {author} {\bibfnamefont
  {R.}~\bibnamefont {Barreiro}},  \emph {et~al.},\ }\href@noop {} {\bibfield
  {journal} {\bibinfo  {journal} {Progress of Theoretical and Experimental
  Physics}\ }\textbf {\bibinfo {volume} {2023}},\ \bibinfo {pages} {042F01}
  (\bibinfo {year} {2023})}\BibitemShut {NoStop}%
\bibitem [{\citenamefont {Eriksen}\ \emph {et~al.}(2007)\citenamefont
  {Eriksen}, \citenamefont {Dickinson}, \citenamefont {Jewell}, \citenamefont
  {Banday}, \citenamefont {G{\'o}rski},\ and\ \citenamefont
  {Lawrence}}]{eriksen2007joint}%
  \BibitemOpen
  \bibfield  {author} {\bibinfo {author} {\bibfnamefont {H.}~\bibnamefont
  {Eriksen}}, \bibinfo {author} {\bibfnamefont {C.}~\bibnamefont {Dickinson}},
  \bibinfo {author} {\bibfnamefont {J.}~\bibnamefont {Jewell}}, \bibinfo
  {author} {\bibfnamefont {A.}~\bibnamefont {Banday}}, \bibinfo {author}
  {\bibfnamefont {K.}~\bibnamefont {G{\'o}rski}}, \ and\ \bibinfo {author}
  {\bibfnamefont {C.}~\bibnamefont {Lawrence}},\ }\href@noop {} {\bibfield
  {journal} {\bibinfo  {journal} {The Astrophysical Journal}\ }\textbf
  {\bibinfo {volume} {672}},\ \bibinfo {pages} {L87} (\bibinfo {year}
  {2007})}\BibitemShut {NoStop}%
\bibitem [{\citenamefont {Eriksen}\ \emph {et~al.}(2008)\citenamefont
  {Eriksen}, \citenamefont {Jewell}, \citenamefont {Dickinson}, \citenamefont
  {Banday}, \citenamefont {G{\'o}rski},\ and\ \citenamefont
  {Lawrence}}]{eriksen2008joint}%
  \BibitemOpen
  \bibfield  {author} {\bibinfo {author} {\bibfnamefont {H.}~\bibnamefont
  {Eriksen}}, \bibinfo {author} {\bibfnamefont {J.}~\bibnamefont {Jewell}},
  \bibinfo {author} {\bibfnamefont {C.}~\bibnamefont {Dickinson}}, \bibinfo
  {author} {\bibfnamefont {A.}~\bibnamefont {Banday}}, \bibinfo {author}
  {\bibfnamefont {K.}~\bibnamefont {G{\'o}rski}}, \ and\ \bibinfo {author}
  {\bibfnamefont {C.}~\bibnamefont {Lawrence}},\ }\href@noop {} {\bibfield
  {journal} {\bibinfo  {journal} {The Astrophysical Journal}\ }\textbf
  {\bibinfo {volume} {676}},\ \bibinfo {pages} {10} (\bibinfo {year}
  {2008})}\BibitemShut {NoStop}%
\bibitem [{\citenamefont {Land}\ and\ \citenamefont
  {Magueijo}(2006)}]{land2006template}%
  \BibitemOpen
  \bibfield  {author} {\bibinfo {author} {\bibfnamefont {K.}~\bibnamefont
  {Land}}\ and\ \bibinfo {author} {\bibfnamefont {J.}~\bibnamefont
  {Magueijo}},\ }\href@noop {} {\bibfield  {journal} {\bibinfo  {journal}
  {Monthly Notices of the Royal Astronomical Society}\ }\textbf {\bibinfo
  {volume} {367}},\ \bibinfo {pages} {1714} (\bibinfo {year}
  {2006})}\BibitemShut {NoStop}%
\bibitem [{\citenamefont {Jaffe}\ \emph {et~al.}(2006)\citenamefont {Jaffe},
  \citenamefont {Banday}, \citenamefont {Eriksen}, \citenamefont {G{\'o}rski},\
  and\ \citenamefont {Hansen}}]{jaffe2006fast}%
  \BibitemOpen
  \bibfield  {author} {\bibinfo {author} {\bibfnamefont {T.}~\bibnamefont
  {Jaffe}}, \bibinfo {author} {\bibfnamefont {A.}~\bibnamefont {Banday}},
  \bibinfo {author} {\bibfnamefont {H.}~\bibnamefont {Eriksen}}, \bibinfo
  {author} {\bibfnamefont {K.}~\bibnamefont {G{\'o}rski}}, \ and\ \bibinfo
  {author} {\bibfnamefont {F.}~\bibnamefont {Hansen}},\ }\href@noop {}
  {\bibfield  {journal} {\bibinfo  {journal} {The Astrophysical Journal}\
  }\textbf {\bibinfo {volume} {643}},\ \bibinfo {pages} {616} (\bibinfo {year}
  {2006})}\BibitemShut {NoStop}%
\bibitem [{\citenamefont {Bennett}\ \emph
  {et~al.}(2003{\natexlab{c}})\citenamefont {Bennett}, \citenamefont {Hill},
  \citenamefont {Hinshaw}, \citenamefont {Nolta}, \citenamefont {Odegard},
  \citenamefont {Page}, \citenamefont {Spergel}, \citenamefont {Weiland},
  \citenamefont {Wright}, \citenamefont {Halpern} \emph
  {et~al.}}]{bennett2003first}%
  \BibitemOpen
  \bibfield  {author} {\bibinfo {author} {\bibfnamefont {C.}~\bibnamefont
  {Bennett}}, \bibinfo {author} {\bibfnamefont {R.~S.}\ \bibnamefont {Hill}},
  \bibinfo {author} {\bibfnamefont {G.}~\bibnamefont {Hinshaw}}, \bibinfo
  {author} {\bibfnamefont {M.}~\bibnamefont {Nolta}}, \bibinfo {author}
  {\bibfnamefont {N.}~\bibnamefont {Odegard}}, \bibinfo {author} {\bibfnamefont
  {L.}~\bibnamefont {Page}}, \bibinfo {author} {\bibfnamefont {D.}~\bibnamefont
  {Spergel}}, \bibinfo {author} {\bibfnamefont {J.}~\bibnamefont {Weiland}},
  \bibinfo {author} {\bibfnamefont {E.}~\bibnamefont {Wright}}, \bibinfo
  {author} {\bibfnamefont {M.}~\bibnamefont {Halpern}},  \emph {et~al.},\
  }\href@noop {} {\bibfield  {journal} {\bibinfo  {journal} {The Astrophysical
  Journal Supplement Series}\ }\textbf {\bibinfo {volume} {148}},\ \bibinfo
  {pages} {97} (\bibinfo {year} {2003}{\natexlab{c}})}\BibitemShut {NoStop}%
\bibitem [{\citenamefont {Eriksen}\ \emph {et~al.}(2004)\citenamefont
  {Eriksen}, \citenamefont {Banday}, \citenamefont {G{\'o}rski},\ and\
  \citenamefont {Lilje}}]{eriksen2004foreground}%
  \BibitemOpen
  \bibfield  {author} {\bibinfo {author} {\bibfnamefont {H.~K.}\ \bibnamefont
  {Eriksen}}, \bibinfo {author} {\bibfnamefont {A.}~\bibnamefont {Banday}},
  \bibinfo {author} {\bibfnamefont {K.}~\bibnamefont {G{\'o}rski}}, \ and\
  \bibinfo {author} {\bibfnamefont {P.}~\bibnamefont {Lilje}},\ }\href@noop {}
  {\bibfield  {journal} {\bibinfo  {journal} {The Astrophysical Journal}\
  }\textbf {\bibinfo {volume} {612}},\ \bibinfo {pages} {633} (\bibinfo {year}
  {2004})}\BibitemShut {NoStop}%
\bibitem [{\citenamefont {Tegmark}\ \emph {et~al.}(2003)\citenamefont
  {Tegmark}, \citenamefont {de~Oliveira-Costa},\ and\ \citenamefont
  {Hamilton}}]{tegmark2003high}%
  \BibitemOpen
  \bibfield  {author} {\bibinfo {author} {\bibfnamefont {M.}~\bibnamefont
  {Tegmark}}, \bibinfo {author} {\bibfnamefont {A.}~\bibnamefont
  {de~Oliveira-Costa}}, \ and\ \bibinfo {author} {\bibfnamefont {A.~J.}\
  \bibnamefont {Hamilton}},\ }\href@noop {} {\bibfield  {journal} {\bibinfo
  {journal} {Physical Review D}\ }\textbf {\bibinfo {volume} {68}},\ \bibinfo
  {pages} {123523} (\bibinfo {year} {2003})}\BibitemShut {NoStop}%
\bibitem [{\citenamefont {Delabrouille}\ \emph {et~al.}(2009)\citenamefont
  {Delabrouille}, \citenamefont {Cardoso}, \citenamefont {Le~Jeune},
  \citenamefont {Betoule}, \citenamefont {Fay},\ and\ \citenamefont
  {Guilloux}}]{delabrouille2009full}%
  \BibitemOpen
  \bibfield  {author} {\bibinfo {author} {\bibfnamefont {J.}~\bibnamefont
  {Delabrouille}}, \bibinfo {author} {\bibfnamefont {J.-F.}\ \bibnamefont
  {Cardoso}}, \bibinfo {author} {\bibfnamefont {M.}~\bibnamefont {Le~Jeune}},
  \bibinfo {author} {\bibfnamefont {M.}~\bibnamefont {Betoule}}, \bibinfo
  {author} {\bibfnamefont {G.}~\bibnamefont {Fay}}, \ and\ \bibinfo {author}
  {\bibfnamefont {F.}~\bibnamefont {Guilloux}},\ }\href@noop {} {\bibfield
  {journal} {\bibinfo  {journal} {Astronomy \& Astrophysics}\ }\textbf
  {\bibinfo {volume} {493}},\ \bibinfo {pages} {835} (\bibinfo {year}
  {2009})}\BibitemShut {NoStop}%
\bibitem [{\citenamefont {Sudevan}\ \emph {et~al.}(2017)\citenamefont
  {Sudevan}, \citenamefont {Aluri}, \citenamefont {Yadav}, \citenamefont
  {Saha},\ and\ \citenamefont {Souradeep}}]{sudevan2017improved}%
  \BibitemOpen
  \bibfield  {author} {\bibinfo {author} {\bibfnamefont {V.}~\bibnamefont
  {Sudevan}}, \bibinfo {author} {\bibfnamefont {P.~K.}\ \bibnamefont {Aluri}},
  \bibinfo {author} {\bibfnamefont {S.~K.}\ \bibnamefont {Yadav}}, \bibinfo
  {author} {\bibfnamefont {R.}~\bibnamefont {Saha}}, \ and\ \bibinfo {author}
  {\bibfnamefont {T.}~\bibnamefont {Souradeep}},\ }\href@noop {} {\bibfield
  {journal} {\bibinfo  {journal} {The Astrophysical Journal}\ }\textbf
  {\bibinfo {volume} {842}},\ \bibinfo {pages} {62} (\bibinfo {year}
  {2017})}\BibitemShut {NoStop}%
\bibitem [{\citenamefont {Sudevan}\ and\ \citenamefont
  {Saha}(2020)}]{Sudevan:2018qyj}%
  \BibitemOpen
  \bibfield  {author} {\bibinfo {author} {\bibfnamefont {V.}~\bibnamefont
  {Sudevan}}\ and\ \bibinfo {author} {\bibfnamefont {R.}~\bibnamefont {Saha}},\
  }\href {\doibase 10.3847/1538-4357/ab964e} {\bibfield  {journal} {\bibinfo
  {journal} {Astrophys. J.}\ }\textbf {\bibinfo {volume} {897}},\ \bibinfo
  {pages} {30} (\bibinfo {year} {2020})},\ \Eprint
  {http://arxiv.org/abs/1810.08872} {arXiv:1810.08872 [astro-ph.CO]}
  \BibitemShut {NoStop}%
\bibitem [{\citenamefont {Sudevan}\ and\ \citenamefont
  {Saha}(2018)}]{Sudevan:2017una}%
  \BibitemOpen
  \bibfield  {author} {\bibinfo {author} {\bibfnamefont {V.}~\bibnamefont
  {Sudevan}}\ and\ \bibinfo {author} {\bibfnamefont {R.}~\bibnamefont {Saha}},\
  }\href {\doibase 10.3847/1538-4357/aae439} {\bibfield  {journal} {\bibinfo
  {journal} {Astrophys. J.}\ }\textbf {\bibinfo {volume} {867}},\ \bibinfo
  {pages} {74} (\bibinfo {year} {2018})},\ \Eprint
  {http://arxiv.org/abs/1712.09804} {arXiv:1712.09804 [astro-ph.CO]}
  \BibitemShut {NoStop}%
\bibitem [{\citenamefont {Taylor}\ \emph {et~al.}(2006)\citenamefont {Taylor},
  \citenamefont {Ashdown},\ and\ \citenamefont {Hobson}}]{Taylor:2006otn}%
  \BibitemOpen
  \bibfield  {author} {\bibinfo {author} {\bibfnamefont {J.~F.}\ \bibnamefont
  {Taylor}}, \bibinfo {author} {\bibfnamefont {M.~A.~J.}\ \bibnamefont
  {Ashdown}}, \ and\ \bibinfo {author} {\bibfnamefont {M.~P.}\ \bibnamefont
  {Hobson}},\ }in\ \href@noop {} {\emph {\bibinfo {booktitle} {{41st Rencontres
  de Moriond: Workshop on Cosmology: Contents and Structures of the
  Universe}}}}\ (\bibinfo  {publisher} {The Gioi},\ \bibinfo {address}
  {Hanoi},\ \bibinfo {year} {2006})\ pp.\ \bibinfo {pages}
  {290--292}\BibitemShut {NoStop}%
\bibitem [{\citenamefont {Hurier}\ \emph {et~al.}(2013)\citenamefont {Hurier},
  \citenamefont {Mac{\'\i}as-P{\'e}rez},\ and\ \citenamefont
  {Hildebrandt}}]{hurier2013milca}%
  \BibitemOpen
  \bibfield  {author} {\bibinfo {author} {\bibfnamefont {G.}~\bibnamefont
  {Hurier}}, \bibinfo {author} {\bibfnamefont {J.}~\bibnamefont
  {Mac{\'\i}as-P{\'e}rez}}, \ and\ \bibinfo {author} {\bibfnamefont
  {S.}~\bibnamefont {Hildebrandt}},\ }\href@noop {} {\bibfield  {journal}
  {\bibinfo  {journal} {Astronomy \& Astrophysics}\ }\textbf {\bibinfo {volume}
  {558}},\ \bibinfo {pages} {A118} (\bibinfo {year} {2013})}\BibitemShut
  {NoStop}%
\bibitem [{\citenamefont {Hivon}\ \emph {et~al.}(2002)\citenamefont {Hivon},
  \citenamefont {G{\'o}rski}, \citenamefont {Netterfield}, \citenamefont
  {Crill}, \citenamefont {Prunet},\ and\ \citenamefont
  {Hansen}}]{hivon2002master}%
  \BibitemOpen
  \bibfield  {author} {\bibinfo {author} {\bibfnamefont {E.}~\bibnamefont
  {Hivon}}, \bibinfo {author} {\bibfnamefont {K.~M.}\ \bibnamefont
  {G{\'o}rski}}, \bibinfo {author} {\bibfnamefont {C.~B.}\ \bibnamefont
  {Netterfield}}, \bibinfo {author} {\bibfnamefont {B.~P.}\ \bibnamefont
  {Crill}}, \bibinfo {author} {\bibfnamefont {S.}~\bibnamefont {Prunet}}, \
  and\ \bibinfo {author} {\bibfnamefont {F.}~\bibnamefont {Hansen}},\
  }\href@noop {} {\bibfield  {journal} {\bibinfo  {journal} {The Astrophysical
  Journal}\ }\textbf {\bibinfo {volume} {567}},\ \bibinfo {pages} {2} (\bibinfo
  {year} {2002})}\BibitemShut {NoStop}%
\bibitem [{\citenamefont {Alonso}\ \emph {et~al.}(2019)\citenamefont {Alonso},
  \citenamefont {Sanchez}, \citenamefont {Slosar},\ and\ \citenamefont
  {Collaboration}}]{alonso2019unified}%
  \BibitemOpen
  \bibfield  {author} {\bibinfo {author} {\bibfnamefont {D.}~\bibnamefont
  {Alonso}}, \bibinfo {author} {\bibfnamefont {J.}~\bibnamefont {Sanchez}},
  \bibinfo {author} {\bibfnamefont {A.}~\bibnamefont {Slosar}}, \ and\ \bibinfo
  {author} {\bibfnamefont {L.~D. E.~S.}\ \bibnamefont {Collaboration}},\
  }\href@noop {} {\bibfield  {journal} {\bibinfo  {journal} {Monthly Notices of
  the Royal Astronomical Society}\ }\textbf {\bibinfo {volume} {484}},\
  \bibinfo {pages} {4127} (\bibinfo {year} {2019})}\BibitemShut {NoStop}%
\bibitem [{\citenamefont {Tegmark}(1997)}]{Tegmark:1996qt}%
  \BibitemOpen
  \bibfield  {author} {\bibinfo {author} {\bibfnamefont {M.}~\bibnamefont
  {Tegmark}},\ }\href {\doibase 10.1103/PhysRevD.55.5895} {\bibfield  {journal}
  {\bibinfo  {journal} {Phys. Rev. D}\ }\textbf {\bibinfo {volume} {55}},\
  \bibinfo {pages} {5895} (\bibinfo {year} {1997})},\ \Eprint
  {http://arxiv.org/abs/astro-ph/9611174} {arXiv:astro-ph/9611174} \BibitemShut
  {NoStop}%
\bibitem [{\citenamefont {Tegmark}\ and\ \citenamefont
  {de~Oliveira-Costa}(2001)}]{Tegmark:2001zv}%
  \BibitemOpen
  \bibfield  {author} {\bibinfo {author} {\bibfnamefont {M.}~\bibnamefont
  {Tegmark}}\ and\ \bibinfo {author} {\bibfnamefont {A.}~\bibnamefont
  {de~Oliveira-Costa}},\ }\href {\doibase 10.1103/PhysRevD.64.063001}
  {\bibfield  {journal} {\bibinfo  {journal} {Phys. Rev. D}\ }\textbf {\bibinfo
  {volume} {64}},\ \bibinfo {pages} {063001} (\bibinfo {year} {2001})},\
  \Eprint {http://arxiv.org/abs/astro-ph/0012120} {arXiv:astro-ph/0012120}
  \BibitemShut {NoStop}%
\bibitem [{\citenamefont {Bilbao-Ahedo}\ \emph {et~al.}(2021)\citenamefont
  {Bilbao-Ahedo}, \citenamefont {Barreiro}, \citenamefont {Vielva},
  \citenamefont {Mart\'\i{}nez-Gonz\'alez},\ and\ \citenamefont
  {Herranz}}]{Bilbao-Ahedo:2021jhn}%
  \BibitemOpen
  \bibfield  {author} {\bibinfo {author} {\bibfnamefont {J.~D.}\ \bibnamefont
  {Bilbao-Ahedo}}, \bibinfo {author} {\bibfnamefont {R.~B.}\ \bibnamefont
  {Barreiro}}, \bibinfo {author} {\bibfnamefont {P.}~\bibnamefont {Vielva}},
  \bibinfo {author} {\bibfnamefont {E.}~\bibnamefont
  {Mart\'\i{}nez-Gonz\'alez}}, \ and\ \bibinfo {author} {\bibfnamefont
  {D.}~\bibnamefont {Herranz}},\ }\href {\doibase
  10.1088/1475-7516/2021/07/034} {\bibfield  {journal} {\bibinfo  {journal}
  {JCAP}\ }\textbf {\bibinfo {volume} {07}},\ \bibinfo {pages} {034} (\bibinfo
  {year} {2021})},\ \Eprint {http://arxiv.org/abs/2104.08528} {arXiv:2104.08528
  [astro-ph.CO]} \BibitemShut {NoStop}%
\bibitem [{\citenamefont {Abrial}\ \emph {et~al.}(2008)\citenamefont {Abrial},
  \citenamefont {Moudden}, \citenamefont {Starck}, \citenamefont {Fadili},
  \citenamefont {Delabrouille},\ and\ \citenamefont {Nguyen}}]{ABRIAL2008289}%
  \BibitemOpen
  \bibfield  {author} {\bibinfo {author} {\bibfnamefont {P.}~\bibnamefont
  {Abrial}}, \bibinfo {author} {\bibfnamefont {Y.}~\bibnamefont {Moudden}},
  \bibinfo {author} {\bibfnamefont {J.-L.}\ \bibnamefont {Starck}}, \bibinfo
  {author} {\bibfnamefont {J.}~\bibnamefont {Fadili}}, \bibinfo {author}
  {\bibfnamefont {J.}~\bibnamefont {Delabrouille}}, \ and\ \bibinfo {author}
  {\bibfnamefont {M.}~\bibnamefont {Nguyen}},\ }\href {\doibase
  https://doi.org/10.1016/j.stamet.2007.11.005} {\bibfield  {journal} {\bibinfo
   {journal} {Statistical Methodology}\ }\textbf {\bibinfo {volume} {5}},\
  \bibinfo {pages} {289} (\bibinfo {year} {2008})},\ \bibinfo {note}
  {astrostatistics}\BibitemShut {NoStop}%
\bibitem [{\citenamefont {{Perotto}}\ \emph {et~al.}(2010)\citenamefont
  {{Perotto}}, \citenamefont {{Bobin}}, \citenamefont {{Plaszczynski}},
  \citenamefont {{Starck}},\ and\ \citenamefont
  {{Lavabre}}}]{2010A&A...519A...4P}%
  \BibitemOpen
  \bibfield  {author} {\bibinfo {author} {\bibfnamefont {L.}~\bibnamefont
  {{Perotto}}}, \bibinfo {author} {\bibfnamefont {J.}~\bibnamefont {{Bobin}}},
  \bibinfo {author} {\bibfnamefont {S.}~\bibnamefont {{Plaszczynski}}},
  \bibinfo {author} {\bibfnamefont {J.~L.}\ \bibnamefont {{Starck}}}, \ and\
  \bibinfo {author} {\bibfnamefont {A.}~\bibnamefont {{Lavabre}}},\ }\href
  {\doibase 10.1051/0004-6361/200912001} {\bibfield  {journal} {\bibinfo
  {journal} {Astronomy \& Astrophysics}\ }\textbf {\bibinfo {volume} {519}},\
  \bibinfo {eid} {A4} (\bibinfo {year} {2010})}\BibitemShut {NoStop}%
\bibitem [{\citenamefont {{Plaszczynski}}\ \emph {et~al.}(2012)\citenamefont
  {{Plaszczynski}}, \citenamefont {{Lavabre}}, \citenamefont {{Perotto}},\ and\
  \citenamefont {{Starck}}}]{2012A&A...544A..27P}%
  \BibitemOpen
  \bibfield  {author} {\bibinfo {author} {\bibfnamefont {S.}~\bibnamefont
  {{Plaszczynski}}}, \bibinfo {author} {\bibfnamefont {A.}~\bibnamefont
  {{Lavabre}}}, \bibinfo {author} {\bibfnamefont {L.}~\bibnamefont
  {{Perotto}}}, \ and\ \bibinfo {author} {\bibfnamefont {J.~L.}\ \bibnamefont
  {{Starck}}},\ }\href {\doibase 10.1051/0004-6361/201218899} {\bibfield
  {journal} {\bibinfo  {journal} {Astronomy \& Astrophysics}\ }\textbf
  {\bibinfo {volume} {544}},\ \bibinfo {eid} {A27} (\bibinfo {year} {2012})},\
  \Eprint {http://arxiv.org/abs/1201.5779} {arXiv:1201.5779 [astro-ph.CO]}
  \BibitemShut {NoStop}%
\bibitem [{\citenamefont {{Dup{\'e}}}\ \emph {et~al.}(2011)\citenamefont
  {{Dup{\'e}}}, \citenamefont {{Rassat}}, \citenamefont {{Starck}},\ and\
  \citenamefont {{Fadili}}}]{2011A&A...534A..51D}%
  \BibitemOpen
  \bibfield  {author} {\bibinfo {author} {\bibfnamefont {F.~X.}\ \bibnamefont
  {{Dup{\'e}}}}, \bibinfo {author} {\bibfnamefont {A.}~\bibnamefont
  {{Rassat}}}, \bibinfo {author} {\bibfnamefont {J.~L.}\ \bibnamefont
  {{Starck}}}, \ and\ \bibinfo {author} {\bibfnamefont {M.~J.}\ \bibnamefont
  {{Fadili}}},\ }\href {\doibase 10.1051/0004-6361/201015893} {\bibfield
  {journal} {\bibinfo  {journal} {Astronomy \& Astrophysics}\ }\textbf
  {\bibinfo {volume} {534}},\ \bibinfo {eid} {A51} (\bibinfo {year} {2011})},\
  \Eprint {http://arxiv.org/abs/1010.2192} {arXiv:1010.2192 [astro-ph.CO]}
  \BibitemShut {NoStop}%
\bibitem [{\citenamefont {Hoffman}\ and\ \citenamefont
  {Ribak}(1991)}]{hoffman1991constrained}%
  \BibitemOpen
  \bibfield  {author} {\bibinfo {author} {\bibfnamefont {Y.}~\bibnamefont
  {Hoffman}}\ and\ \bibinfo {author} {\bibfnamefont {E.}~\bibnamefont
  {Ribak}},\ }\href@noop {} {\bibfield  {journal} {\bibinfo  {journal}
  {Astrophysical Journal, Part 2-Letters (ISSN 0004-637X), vol. 380, Oct. 10,
  1991, p. L5-L8.}\ }\textbf {\bibinfo {volume} {380}},\ \bibinfo {pages} {L5}
  (\bibinfo {year} {1991})}\BibitemShut {NoStop}%
\bibitem [{\citenamefont {Bucher}\ and\ \citenamefont
  {Louis}(2012)}]{bucher2012filling}%
  \BibitemOpen
  \bibfield  {author} {\bibinfo {author} {\bibfnamefont {M.}~\bibnamefont
  {Bucher}}\ and\ \bibinfo {author} {\bibfnamefont {T.}~\bibnamefont {Louis}},\
  }\href@noop {} {\bibfield  {journal} {\bibinfo  {journal} {Monthly Notices of
  the Royal Astronomical Society}\ }\textbf {\bibinfo {volume} {424}},\
  \bibinfo {pages} {1694} (\bibinfo {year} {2012})}\BibitemShut {NoStop}%
\bibitem [{\citenamefont {{Kim}}\ \emph {et~al.}(2012)\citenamefont {{Kim}},
  \citenamefont {{Naselsky}},\ and\ \citenamefont
  {{Mandolesi}}}]{2012ApJ...750L...9K}%
  \BibitemOpen
  \bibfield  {author} {\bibinfo {author} {\bibfnamefont {J.}~\bibnamefont
  {{Kim}}}, \bibinfo {author} {\bibfnamefont {P.}~\bibnamefont {{Naselsky}}}, \
  and\ \bibinfo {author} {\bibfnamefont {N.}~\bibnamefont {{Mandolesi}}},\
  }\href {\doibase 10.1088/2041-8205/750/1/L9} {\bibfield  {journal} {\bibinfo
  {journal} {The Astrophysical Journal Letters}\ }\textbf {\bibinfo {volume}
  {750}},\ \bibinfo {eid} {L9} (\bibinfo {year} {2012})},\ \Eprint
  {http://arxiv.org/abs/1202.0188} {arXiv:1202.0188 [astro-ph.CO]} \BibitemShut
  {NoStop}%
\bibitem [{\citenamefont {{Feeney}}\ \emph {et~al.}(2011)\citenamefont
  {{Feeney}}, \citenamefont {{Peiris}},\ and\ \citenamefont
  {{Pontzen}}}]{2011PhRvD..84j3002F}%
  \BibitemOpen
  \bibfield  {author} {\bibinfo {author} {\bibfnamefont {S.~M.}\ \bibnamefont
  {{Feeney}}}, \bibinfo {author} {\bibfnamefont {H.~V.}\ \bibnamefont
  {{Peiris}}}, \ and\ \bibinfo {author} {\bibfnamefont {A.}~\bibnamefont
  {{Pontzen}}},\ }\href {\doibase 10.1103/PhysRevD.84.103002} {\bibfield
  {journal} {\bibinfo  {journal} {Physical Review D}\ }\textbf {\bibinfo
  {volume} {84}},\ \bibinfo {eid} {103002} (\bibinfo {year} {2011})},\ \Eprint
  {http://arxiv.org/abs/1107.5466} {arXiv:1107.5466 [astro-ph.CO]} \BibitemShut
  {NoStop}%
\bibitem [{\citenamefont {Copi}\ \emph {et~al.}(2011)\citenamefont {Copi},
  \citenamefont {Huterer}, \citenamefont {Schwarz},\ and\ \citenamefont
  {Starkman}}]{copi2011bias}%
  \BibitemOpen
  \bibfield  {author} {\bibinfo {author} {\bibfnamefont {C.~J.}\ \bibnamefont
  {Copi}}, \bibinfo {author} {\bibfnamefont {D.}~\bibnamefont {Huterer}},
  \bibinfo {author} {\bibfnamefont {D.~J.}\ \bibnamefont {Schwarz}}, \ and\
  \bibinfo {author} {\bibfnamefont {G.~D.}\ \bibnamefont {Starkman}},\
  }\href@noop {} {\bibfield  {journal} {\bibinfo  {journal} {Monthly Notices of
  the Royal Astronomical Society}\ }\textbf {\bibinfo {volume} {418}},\
  \bibinfo {pages} {505} (\bibinfo {year} {2011})}\BibitemShut {NoStop}%
\bibitem [{\citenamefont {Ade}\ \emph {et~al.}(2014{\natexlab{b}})\citenamefont
  {Ade} \emph {et~al.}}]{Planck:2013wtn}%
  \BibitemOpen
  \bibfield  {author} {\bibinfo {author} {\bibfnamefont {P.~A.~R.}\
  \bibnamefont {Ade}} \emph {et~al.} (\bibinfo {collaboration} {Planck}),\
  }\href {\doibase 10.1051/0004-6361/201321554} {\bibfield  {journal} {\bibinfo
   {journal} {Astron. Astrophys.}\ }\textbf {\bibinfo {volume} {571}},\
  \bibinfo {pages} {A24} (\bibinfo {year} {2014}{\natexlab{b}})},\ \Eprint
  {http://arxiv.org/abs/1303.5084} {arXiv:1303.5084 [astro-ph.CO]} \BibitemShut
  {NoStop}%
\bibitem [{\citenamefont {Gimeno-Amo}\ \emph {et~al.}(2024)\citenamefont
  {Gimeno-Amo}, \citenamefont {Mart\'\i{}nez-Gonz\'alez},\ and\ \citenamefont
  {Barreiro}}]{Gimeno-Amo:2024hca}%
  \BibitemOpen
  \bibfield  {author} {\bibinfo {author} {\bibfnamefont {C.}~\bibnamefont
  {Gimeno-Amo}}, \bibinfo {author} {\bibfnamefont {E.}~\bibnamefont
  {Mart\'\i{}nez-Gonz\'alez}}, \ and\ \bibinfo {author} {\bibfnamefont {R.~B.}\
  \bibnamefont {Barreiro}},\ }\href {\doibase 10.1088/1475-7516/2024/09/038}
  {\bibfield  {journal} {\bibinfo  {journal} {JCAP}\ }\textbf {\bibinfo
  {volume} {09}},\ \bibinfo {pages} {038} (\bibinfo {year} {2024})},\ \Eprint
  {http://arxiv.org/abs/2405.06820} {arXiv:2405.06820 [astro-ph.CO]}
  \BibitemShut {NoStop}%
\bibitem [{\citenamefont {Marcos-Caballero}\ and\ \citenamefont
  {Mart\'\i{}nez-Gonz\'alez}(2019)}]{Marcos-Caballero:2019jqj}%
  \BibitemOpen
  \bibfield  {author} {\bibinfo {author} {\bibfnamefont {A.}~\bibnamefont
  {Marcos-Caballero}}\ and\ \bibinfo {author} {\bibfnamefont {E.}~\bibnamefont
  {Mart\'\i{}nez-Gonz\'alez}},\ }\href {\doibase 10.1088/1475-7516/2019/10/053}
  {\bibfield  {journal} {\bibinfo  {journal} {JCAP}\ }\textbf {\bibinfo
  {volume} {10}},\ \bibinfo {pages} {053} (\bibinfo {year} {2019})},\ \Eprint
  {http://arxiv.org/abs/1909.06093} {arXiv:1909.06093 [astro-ph.CO]}
  \BibitemShut {NoStop}%
\bibitem [{\citenamefont {Hochreiter}(1997)}]{hochreiter1997long}%
  \BibitemOpen
  \bibfield  {author} {\bibinfo {author} {\bibfnamefont {S.}~\bibnamefont
  {Hochreiter}},\ }\href@noop {} {\bibfield  {journal} {\bibinfo  {journal}
  {Neural Computation MIT-Press}\ } (\bibinfo {year} {1997})}\BibitemShut
  {NoStop}%
\bibitem [{\citenamefont {Goodfellow}\ \emph {et~al.}(2014)\citenamefont
  {Goodfellow}, \citenamefont {Pouget-Abadie}, \citenamefont {Mirza},
  \citenamefont {Xu}, \citenamefont {Warde-Farley}, \citenamefont {Ozair},
  \citenamefont {Courville},\ and\ \citenamefont
  {Bengio}}]{goodfellow2014generative}%
  \BibitemOpen
  \bibfield  {author} {\bibinfo {author} {\bibfnamefont {I.}~\bibnamefont
  {Goodfellow}}, \bibinfo {author} {\bibfnamefont {J.}~\bibnamefont
  {Pouget-Abadie}}, \bibinfo {author} {\bibfnamefont {M.}~\bibnamefont
  {Mirza}}, \bibinfo {author} {\bibfnamefont {B.}~\bibnamefont {Xu}}, \bibinfo
  {author} {\bibfnamefont {D.}~\bibnamefont {Warde-Farley}}, \bibinfo {author}
  {\bibfnamefont {S.}~\bibnamefont {Ozair}}, \bibinfo {author} {\bibfnamefont
  {A.}~\bibnamefont {Courville}}, \ and\ \bibinfo {author} {\bibfnamefont
  {Y.}~\bibnamefont {Bengio}},\ }\href@noop {} {\bibfield  {journal} {\bibinfo
  {journal} {Advances in neural information processing systems}\ }\textbf
  {\bibinfo {volume} {27}} (\bibinfo {year} {2014})}\BibitemShut {NoStop}%
\bibitem [{\citenamefont {Vaswani}(2017)}]{vaswani2017attention}%
  \BibitemOpen
  \bibfield  {author} {\bibinfo {author} {\bibfnamefont {A.}~\bibnamefont
  {Vaswani}},\ }\href@noop {} {\bibfield  {journal} {\bibinfo  {journal}
  {Advances in Neural Information Processing Systems}\ } (\bibinfo {year}
  {2017})}\BibitemShut {NoStop}%
\bibitem [{\citenamefont {Krizhevsky}\ \emph {et~al.}(2017)\citenamefont
  {Krizhevsky}, \citenamefont {Sutskever},\ and\ \citenamefont
  {Hinton}}]{krizhevsky2017imagenet}%
  \BibitemOpen
  \bibfield  {author} {\bibinfo {author} {\bibfnamefont {A.}~\bibnamefont
  {Krizhevsky}}, \bibinfo {author} {\bibfnamefont {I.}~\bibnamefont
  {Sutskever}}, \ and\ \bibinfo {author} {\bibfnamefont {G.~E.}\ \bibnamefont
  {Hinton}},\ }\href@noop {} {\bibfield  {journal} {\bibinfo  {journal}
  {Communications of the ACM}\ }\textbf {\bibinfo {volume} {60}},\ \bibinfo
  {pages} {84} (\bibinfo {year} {2017})}\BibitemShut {NoStop}%
\bibitem [{\citenamefont {Dai}\ \emph {et~al.}(2021)\citenamefont {Dai},
  \citenamefont {Liu}, \citenamefont {Le},\ and\ \citenamefont
  {Tan}}]{dai2021coatnet}%
  \BibitemOpen
  \bibfield  {author} {\bibinfo {author} {\bibfnamefont {Z.}~\bibnamefont
  {Dai}}, \bibinfo {author} {\bibfnamefont {H.}~\bibnamefont {Liu}}, \bibinfo
  {author} {\bibfnamefont {Q.~V.}\ \bibnamefont {Le}}, \ and\ \bibinfo {author}
  {\bibfnamefont {M.}~\bibnamefont {Tan}},\ }\href@noop {} {\bibfield
  {journal} {\bibinfo  {journal} {Advances in neural information processing
  systems}\ }\textbf {\bibinfo {volume} {34}},\ \bibinfo {pages} {3965}
  (\bibinfo {year} {2021})}\BibitemShut {NoStop}%
\bibitem [{\citenamefont {Cai}\ \emph {et~al.}(2022)\citenamefont {Cai},
  \citenamefont {Zhou}, \citenamefont {Han}, \citenamefont {Sun}, \citenamefont
  {Kong}, \citenamefont {Li},\ and\ \citenamefont {Zhang}}]{cai2022reversible}%
  \BibitemOpen
  \bibfield  {author} {\bibinfo {author} {\bibfnamefont {Y.}~\bibnamefont
  {Cai}}, \bibinfo {author} {\bibfnamefont {Y.}~\bibnamefont {Zhou}}, \bibinfo
  {author} {\bibfnamefont {Q.}~\bibnamefont {Han}}, \bibinfo {author}
  {\bibfnamefont {J.}~\bibnamefont {Sun}}, \bibinfo {author} {\bibfnamefont
  {X.}~\bibnamefont {Kong}}, \bibinfo {author} {\bibfnamefont {J.}~\bibnamefont
  {Li}}, \ and\ \bibinfo {author} {\bibfnamefont {X.}~\bibnamefont {Zhang}},\
  }\href@noop {} {\bibfield  {journal} {\bibinfo  {journal} {arXiv preprint
  arXiv:2212.11696}\ } (\bibinfo {year} {2022})}\BibitemShut {NoStop}%
\bibitem [{\citenamefont {Zhou}\ \emph {et~al.}(2021)\citenamefont {Zhou},
  \citenamefont {Koltun},\ and\ \citenamefont
  {Kr{\"a}henb{\"u}hl}}]{zhou2021probabilistic}%
  \BibitemOpen
  \bibfield  {author} {\bibinfo {author} {\bibfnamefont {X.}~\bibnamefont
  {Zhou}}, \bibinfo {author} {\bibfnamefont {V.}~\bibnamefont {Koltun}}, \ and\
  \bibinfo {author} {\bibfnamefont {P.}~\bibnamefont {Kr{\"a}henb{\"u}hl}},\
  }\href@noop {} {\bibfield  {journal} {\bibinfo  {journal} {arXiv preprint
  arXiv:2103.07461}\ } (\bibinfo {year} {2021})}\BibitemShut {NoStop}%
\bibitem [{\citenamefont {Maaz}\ \emph {et~al.}(2022)\citenamefont {Maaz},
  \citenamefont {Shaker}, \citenamefont {Cholakkal}, \citenamefont {Khan},
  \citenamefont {Zamir}, \citenamefont {Anwer},\ and\ \citenamefont
  {Shahbaz~Khan}}]{maaz2022edgenext}%
  \BibitemOpen
  \bibfield  {author} {\bibinfo {author} {\bibfnamefont {M.}~\bibnamefont
  {Maaz}}, \bibinfo {author} {\bibfnamefont {A.}~\bibnamefont {Shaker}},
  \bibinfo {author} {\bibfnamefont {H.}~\bibnamefont {Cholakkal}}, \bibinfo
  {author} {\bibfnamefont {S.}~\bibnamefont {Khan}}, \bibinfo {author}
  {\bibfnamefont {S.~W.}\ \bibnamefont {Zamir}}, \bibinfo {author}
  {\bibfnamefont {R.~M.}\ \bibnamefont {Anwer}}, \ and\ \bibinfo {author}
  {\bibfnamefont {F.}~\bibnamefont {Shahbaz~Khan}},\ }in\ \href@noop {} {\emph
  {\bibinfo {booktitle} {European conference on computer vision}}}\ (\bibinfo
  {organization} {Springer},\ \bibinfo {year} {2022})\ pp.\ \bibinfo {pages}
  {3--20}\BibitemShut {NoStop}%
\bibitem [{\citenamefont {Dai}\ \emph {et~al.}(2017)\citenamefont {Dai},
  \citenamefont {Qi}, \citenamefont {Xiong}, \citenamefont {Li}, \citenamefont
  {Zhang}, \citenamefont {Hu},\ and\ \citenamefont {Wei}}]{dai2017deformable}%
  \BibitemOpen
  \bibfield  {author} {\bibinfo {author} {\bibfnamefont {J.}~\bibnamefont
  {Dai}}, \bibinfo {author} {\bibfnamefont {H.}~\bibnamefont {Qi}}, \bibinfo
  {author} {\bibfnamefont {Y.}~\bibnamefont {Xiong}}, \bibinfo {author}
  {\bibfnamefont {Y.}~\bibnamefont {Li}}, \bibinfo {author} {\bibfnamefont
  {G.}~\bibnamefont {Zhang}}, \bibinfo {author} {\bibfnamefont
  {H.}~\bibnamefont {Hu}}, \ and\ \bibinfo {author} {\bibfnamefont
  {Y.}~\bibnamefont {Wei}},\ }in\ \href@noop {} {\emph {\bibinfo {booktitle}
  {Proceedings of the IEEE international conference on computer vision}}}\
  (\bibinfo {year} {2017})\ pp.\ \bibinfo {pages} {764--773}\BibitemShut
  {NoStop}%
\bibitem [{\citenamefont {Xie}\ \emph {et~al.}(2012)\citenamefont {Xie},
  \citenamefont {Xu},\ and\ \citenamefont {Chen}}]{NIPS2012_6cdd60ea}%
  \BibitemOpen
  \bibfield  {author} {\bibinfo {author} {\bibfnamefont {J.}~\bibnamefont
  {Xie}}, \bibinfo {author} {\bibfnamefont {L.}~\bibnamefont {Xu}}, \ and\
  \bibinfo {author} {\bibfnamefont {E.}~\bibnamefont {Chen}},\ }in\ \href
  {https://proceedings.neurips.cc/paper_files/paper/2012/file/6cdd60ea0045eb7a6ec44c54d29ed402-Paper.pdf}
  {\emph {\bibinfo {booktitle} {Advances in Neural Information Processing
  Systems}}},\ Vol.~\bibinfo {volume} {25},\ \bibinfo {editor} {edited by\
  \bibinfo {editor} {\bibfnamefont {F.}~\bibnamefont {Pereira}}, \bibinfo
  {editor} {\bibfnamefont {C.}~\bibnamefont {Burges}}, \bibinfo {editor}
  {\bibfnamefont {L.}~\bibnamefont {Bottou}}, \ and\ \bibinfo {editor}
  {\bibfnamefont {K.}~\bibnamefont {Weinberger}}}\ (\bibinfo  {publisher}
  {Curran Associates, Inc.},\ \bibinfo {year} {2012})\BibitemShut {NoStop}%
\bibitem [{\citenamefont {Sohl-Dickstein}\ \emph {et~al.}(2015)\citenamefont
  {Sohl-Dickstein}, \citenamefont {Weiss}, \citenamefont {Maheswaranathan},\
  and\ \citenamefont {Ganguli}}]{pmlr-v37-sohl-dickstein15}%
  \BibitemOpen
  \bibfield  {author} {\bibinfo {author} {\bibfnamefont {J.}~\bibnamefont
  {Sohl-Dickstein}}, \bibinfo {author} {\bibfnamefont {E.}~\bibnamefont
  {Weiss}}, \bibinfo {author} {\bibfnamefont {N.}~\bibnamefont
  {Maheswaranathan}}, \ and\ \bibinfo {author} {\bibfnamefont {S.}~\bibnamefont
  {Ganguli}},\ }in\ \href
  {https://proceedings.mlr.press/v37/sohl-dickstein15.html} {\emph {\bibinfo
  {booktitle} {Proceedings of the 32nd International Conference on Machine
  Learning}}},\ \bibinfo {series} {Proceedings of Machine Learning Research},
  Vol.~\bibinfo {volume} {37},\ \bibinfo {editor} {edited by\ \bibinfo {editor}
  {\bibfnamefont {F.}~\bibnamefont {Bach}}\ and\ \bibinfo {editor}
  {\bibfnamefont {D.}~\bibnamefont {Blei}}}\ (\bibinfo  {publisher} {PMLR},\
  \bibinfo {address} {Lille, France},\ \bibinfo {year} {2015})\ pp.\ \bibinfo
  {pages} {2256--2265}\BibitemShut {NoStop}%
\bibitem [{\citenamefont {Ren}\ \emph {et~al.}(2015)\citenamefont {Ren},
  \citenamefont {Xu}, \citenamefont {Yan},\ and\ \citenamefont
  {Sun}}]{NIPS2015_daca4121}%
  \BibitemOpen
  \bibfield  {author} {\bibinfo {author} {\bibfnamefont {J.~S.}\ \bibnamefont
  {Ren}}, \bibinfo {author} {\bibfnamefont {L.}~\bibnamefont {Xu}}, \bibinfo
  {author} {\bibfnamefont {Q.}~\bibnamefont {Yan}}, \ and\ \bibinfo {author}
  {\bibfnamefont {W.}~\bibnamefont {Sun}},\ }in\ \href
  {https://proceedings.neurips.cc/paper_files/paper/2015/file/daca41214b39c5dc66674d09081940f0-Paper.pdf}
  {\emph {\bibinfo {booktitle} {Advances in Neural Information Processing
  Systems}}},\ Vol.~\bibinfo {volume} {28},\ \bibinfo {editor} {edited by\
  \bibinfo {editor} {\bibfnamefont {C.}~\bibnamefont {Cortes}}, \bibinfo
  {editor} {\bibfnamefont {N.}~\bibnamefont {Lawrence}}, \bibinfo {editor}
  {\bibfnamefont {D.}~\bibnamefont {Lee}}, \bibinfo {editor} {\bibfnamefont
  {M.}~\bibnamefont {Sugiyama}}, \ and\ \bibinfo {editor} {\bibfnamefont
  {R.}~\bibnamefont {Garnett}}}\ (\bibinfo  {publisher} {Curran Associates,
  Inc.},\ \bibinfo {year} {2015})\BibitemShut {NoStop}%
\bibitem [{\citenamefont {Cai}\ \emph {et~al.}(2017)\citenamefont {Cai},
  \citenamefont {Su}, \citenamefont {Lin}, \citenamefont {Wang}, \citenamefont
  {Yang},\ and\ \citenamefont {Ling}}]{blindinpaint}%
  \BibitemOpen
  \bibfield  {author} {\bibinfo {author} {\bibfnamefont {N.}~\bibnamefont
  {Cai}}, \bibinfo {author} {\bibfnamefont {Z.}~\bibnamefont {Su}}, \bibinfo
  {author} {\bibfnamefont {Z.}~\bibnamefont {Lin}}, \bibinfo {author}
  {\bibfnamefont {H.}~\bibnamefont {Wang}}, \bibinfo {author} {\bibfnamefont
  {Z.}~\bibnamefont {Yang}}, \ and\ \bibinfo {author} {\bibfnamefont
  {B.}~\bibnamefont {Ling}},\ }\href {\doibase 10.1007/s00371-015-1190-z}
  {\bibfield  {journal} {\bibinfo  {journal} {The Visual Computer}\ }\textbf
  {\bibinfo {volume} {33}} (\bibinfo {year} {2017}),\
  10.1007/s00371-015-1190-z}\BibitemShut {NoStop}%
\bibitem [{\citenamefont {{Pathak}}\ \emph {et~al.}(2016)\citenamefont
  {{Pathak}}, \citenamefont {{Krahenbuhl}}, \citenamefont {{Donahue}},
  \citenamefont {{Darrell}},\ and\ \citenamefont
  {{Efros}}}]{2016arXiv160407379P}%
  \BibitemOpen
  \bibfield  {author} {\bibinfo {author} {\bibfnamefont {D.}~\bibnamefont
  {{Pathak}}}, \bibinfo {author} {\bibfnamefont {P.}~\bibnamefont
  {{Krahenbuhl}}}, \bibinfo {author} {\bibfnamefont {J.}~\bibnamefont
  {{Donahue}}}, \bibinfo {author} {\bibfnamefont {T.}~\bibnamefont
  {{Darrell}}}, \ and\ \bibinfo {author} {\bibfnamefont {A.~A.}\ \bibnamefont
  {{Efros}}},\ }\href {\doibase 10.48550/arXiv.1604.07379} {\bibfield
  {journal} {\bibinfo  {journal} {arXiv e-prints}\ ,\ \bibinfo {eid}
  {arXiv:1604.07379}} (\bibinfo {year} {2016})},\ \Eprint
  {http://arxiv.org/abs/1604.07379} {arXiv:1604.07379 [cs.CV]} \BibitemShut
  {NoStop}%
\bibitem [{\citenamefont {Petroff}\ \emph {et~al.}(2020)\citenamefont
  {Petroff}, \citenamefont {Addison}, \citenamefont {Bennett},\ and\
  \citenamefont {Weiland}}]{petroff2020full}%
  \BibitemOpen
  \bibfield  {author} {\bibinfo {author} {\bibfnamefont {M.~A.}\ \bibnamefont
  {Petroff}}, \bibinfo {author} {\bibfnamefont {G.~E.}\ \bibnamefont
  {Addison}}, \bibinfo {author} {\bibfnamefont {C.~L.}\ \bibnamefont
  {Bennett}}, \ and\ \bibinfo {author} {\bibfnamefont {J.~L.}\ \bibnamefont
  {Weiland}},\ }\href@noop {} {\bibfield  {journal} {\bibinfo  {journal} {The
  Astrophysical Journal}\ }\textbf {\bibinfo {volume} {903}},\ \bibinfo {pages}
  {104} (\bibinfo {year} {2020})}\BibitemShut {NoStop}%
\bibitem [{\citenamefont {Wang}\ \emph {et~al.}(2022)\citenamefont {Wang},
  \citenamefont {Shi}, \citenamefont {Yan}, \citenamefont {Xia}, \citenamefont
  {Zhao}, \citenamefont {Li},\ and\ \citenamefont {Li}}]{wang2022recovering}%
  \BibitemOpen
  \bibfield  {author} {\bibinfo {author} {\bibfnamefont {G.-J.}\ \bibnamefont
  {Wang}}, \bibinfo {author} {\bibfnamefont {H.-L.}\ \bibnamefont {Shi}},
  \bibinfo {author} {\bibfnamefont {Y.-P.}\ \bibnamefont {Yan}}, \bibinfo
  {author} {\bibfnamefont {J.-Q.}\ \bibnamefont {Xia}}, \bibinfo {author}
  {\bibfnamefont {Y.-Y.}\ \bibnamefont {Zhao}}, \bibinfo {author}
  {\bibfnamefont {S.-Y.}\ \bibnamefont {Li}}, \ and\ \bibinfo {author}
  {\bibfnamefont {J.-F.}\ \bibnamefont {Li}},\ }\href@noop {} {\bibfield
  {journal} {\bibinfo  {journal} {The Astrophysical Journal Supplement Series}\
  }\textbf {\bibinfo {volume} {260}},\ \bibinfo {pages} {13} (\bibinfo {year}
  {2022})}\BibitemShut {NoStop}%
\bibitem [{\citenamefont {Casas}\ \emph {et~al.}(2022)\citenamefont {Casas},
  \citenamefont {Bonavera}, \citenamefont {Gonz{\'a}lez-Nuevo}, \citenamefont
  {Baccigalupi}, \citenamefont {Cueli}, \citenamefont {Crespo}, \citenamefont
  {Goitia}, \citenamefont {Santos}, \citenamefont {S{\'a}nchez},\ and\
  \citenamefont {de~Cos}}]{casas2022cenn}%
  \BibitemOpen
  \bibfield  {author} {\bibinfo {author} {\bibfnamefont {J.}~\bibnamefont
  {Casas}}, \bibinfo {author} {\bibfnamefont {L.}~\bibnamefont {Bonavera}},
  \bibinfo {author} {\bibfnamefont {J.}~\bibnamefont {Gonz{\'a}lez-Nuevo}},
  \bibinfo {author} {\bibfnamefont {C.}~\bibnamefont {Baccigalupi}}, \bibinfo
  {author} {\bibfnamefont {M.}~\bibnamefont {Cueli}}, \bibinfo {author}
  {\bibfnamefont {D.}~\bibnamefont {Crespo}}, \bibinfo {author} {\bibfnamefont
  {E.}~\bibnamefont {Goitia}}, \bibinfo {author} {\bibfnamefont
  {J.}~\bibnamefont {Santos}}, \bibinfo {author} {\bibfnamefont
  {M.}~\bibnamefont {S{\'a}nchez}}, \ and\ \bibinfo {author} {\bibfnamefont
  {F.}~\bibnamefont {de~Cos}},\ }\href@noop {} {\bibfield  {journal} {\bibinfo
  {journal} {Astronomy \& Astrophysics}\ }\textbf {\bibinfo {volume} {666}},\
  \bibinfo {pages} {A89} (\bibinfo {year} {2022})}\BibitemShut {NoStop}%
\bibitem [{\citenamefont {Yan}\ \emph {et~al.}(2023{\natexlab{a}})\citenamefont
  {Yan}, \citenamefont {Wang}, \citenamefont {Li},\ and\ \citenamefont
  {Xia}}]{yan2023recovering}%
  \BibitemOpen
  \bibfield  {author} {\bibinfo {author} {\bibfnamefont {Y.-P.}\ \bibnamefont
  {Yan}}, \bibinfo {author} {\bibfnamefont {G.-J.}\ \bibnamefont {Wang}},
  \bibinfo {author} {\bibfnamefont {S.-Y.}\ \bibnamefont {Li}}, \ and\ \bibinfo
  {author} {\bibfnamefont {J.-Q.}\ \bibnamefont {Xia}},\ }\href@noop {}
  {\bibfield  {journal} {\bibinfo  {journal} {The Astrophysical Journal}\
  }\textbf {\bibinfo {volume} {947}},\ \bibinfo {pages} {29} (\bibinfo {year}
  {2023}{\natexlab{a}})}\BibitemShut {NoStop}%
\bibitem [{\citenamefont {Yan}\ \emph {et~al.}(2024)\citenamefont {Yan},
  \citenamefont {Li}, \citenamefont {Wang}, \citenamefont {Zhang},\ and\
  \citenamefont {Xia}}]{yan2024cmbfscnn}%
  \BibitemOpen
  \bibfield  {author} {\bibinfo {author} {\bibfnamefont {Y.-P.}\ \bibnamefont
  {Yan}}, \bibinfo {author} {\bibfnamefont {S.-Y.}\ \bibnamefont {Li}},
  \bibinfo {author} {\bibfnamefont {G.-J.}\ \bibnamefont {Wang}}, \bibinfo
  {author} {\bibfnamefont {Z.}~\bibnamefont {Zhang}}, \ and\ \bibinfo {author}
  {\bibfnamefont {J.-Q.}\ \bibnamefont {Xia}},\ }\href@noop {} {\bibfield
  {journal} {\bibinfo  {journal} {The Astrophysical Journal Supplement Series}\
  }\textbf {\bibinfo {volume} {274}},\ \bibinfo {pages} {4} (\bibinfo {year}
  {2024})}\BibitemShut {NoStop}%
\bibitem [{\citenamefont {Sudevan}\ and\ \citenamefont
  {Chen}(2024)}]{Sudevan:2024hwq}%
  \BibitemOpen
  \bibfield  {author} {\bibinfo {author} {\bibfnamefont {V.}~\bibnamefont
  {Sudevan}}\ and\ \bibinfo {author} {\bibfnamefont {P.}~\bibnamefont {Chen}},\
  }\href@noop {} {\  (\bibinfo {year} {2024})},\ \Eprint
  {http://arxiv.org/abs/2406.19367} {arXiv:2406.19367 [astro-ph.CO]}
  \BibitemShut {NoStop}%
\bibitem [{\citenamefont {Yan}\ \emph {et~al.}(2023{\natexlab{b}})\citenamefont
  {Yan}, \citenamefont {Wang}, \citenamefont {Li},\ and\ \citenamefont
  {Xia}}]{yan2023delensing}%
  \BibitemOpen
  \bibfield  {author} {\bibinfo {author} {\bibfnamefont {Y.-P.}\ \bibnamefont
  {Yan}}, \bibinfo {author} {\bibfnamefont {G.-J.}\ \bibnamefont {Wang}},
  \bibinfo {author} {\bibfnamefont {S.-Y.}\ \bibnamefont {Li}}, \ and\ \bibinfo
  {author} {\bibfnamefont {J.-Q.}\ \bibnamefont {Xia}},\ }\href@noop {}
  {\bibfield  {journal} {\bibinfo  {journal} {The Astrophysical Journal
  Supplement Series}\ }\textbf {\bibinfo {volume} {267}},\ \bibinfo {pages} {2}
  (\bibinfo {year} {2023}{\natexlab{b}})}\BibitemShut {NoStop}%
\bibitem [{\citenamefont {Puglisi}\ and\ \citenamefont
  {Bai}(2020)}]{puglisi2020inpainting}%
  \BibitemOpen
  \bibfield  {author} {\bibinfo {author} {\bibfnamefont {G.}~\bibnamefont
  {Puglisi}}\ and\ \bibinfo {author} {\bibfnamefont {X.}~\bibnamefont {Bai}},\
  }\href@noop {} {\bibfield  {journal} {\bibinfo  {journal} {The Astrophysical
  Journal}\ }\textbf {\bibinfo {volume} {905}},\ \bibinfo {pages} {143}
  (\bibinfo {year} {2020})}\BibitemShut {NoStop}%
\bibitem [{\citenamefont {Yi}\ \emph {et~al.}(2020)\citenamefont {Yi},
  \citenamefont {Guo}, \citenamefont {Fan}, \citenamefont {Hamann},\ and\
  \citenamefont {Wang}}]{yi2020cosmo}%
  \BibitemOpen
  \bibfield  {author} {\bibinfo {author} {\bibfnamefont {K.}~\bibnamefont
  {Yi}}, \bibinfo {author} {\bibfnamefont {Y.}~\bibnamefont {Guo}}, \bibinfo
  {author} {\bibfnamefont {Y.}~\bibnamefont {Fan}}, \bibinfo {author}
  {\bibfnamefont {J.}~\bibnamefont {Hamann}}, \ and\ \bibinfo {author}
  {\bibfnamefont {Y.~G.}\ \bibnamefont {Wang}},\ }in\ \href@noop {} {\emph
  {\bibinfo {booktitle} {2020 International Joint Conference on Neural Networks
  (IJCNN)}}}\ (\bibinfo {organization} {IEEE},\ \bibinfo {year} {2020})\ pp.\
  \bibinfo {pages} {1--7}\BibitemShut {NoStop}%
\bibitem [{\citenamefont {Sadr}\ and\ \citenamefont
  {Farsian}(2021)}]{Sadr:2020rje}%
  \BibitemOpen
  \bibfield  {author} {\bibinfo {author} {\bibfnamefont {A.~V.}\ \bibnamefont
  {Sadr}}\ and\ \bibinfo {author} {\bibfnamefont {F.}~\bibnamefont {Farsian}},\
  }\href {\doibase 10.1088/1475-7516/2021/03/012} {\bibfield  {journal}
  {\bibinfo  {journal} {JCAP}\ }\textbf {\bibinfo {volume} {03}},\ \bibinfo
  {pages} {012} (\bibinfo {year} {2021})},\ \Eprint
  {http://arxiv.org/abs/2004.04177} {arXiv:2004.04177 [astro-ph.CO]}
  \BibitemShut {NoStop}%
\bibitem [{\citenamefont {Montefalcone}\ \emph {et~al.}(2021)\citenamefont
  {Montefalcone}, \citenamefont {Abitbol}, \citenamefont {Kodwani},\ and\
  \citenamefont {Grumitt}}]{montefalcone2021inpainting}%
  \BibitemOpen
  \bibfield  {author} {\bibinfo {author} {\bibfnamefont {G.}~\bibnamefont
  {Montefalcone}}, \bibinfo {author} {\bibfnamefont {M.~H.}\ \bibnamefont
  {Abitbol}}, \bibinfo {author} {\bibfnamefont {D.}~\bibnamefont {Kodwani}}, \
  and\ \bibinfo {author} {\bibfnamefont {R.}~\bibnamefont {Grumitt}},\
  }\href@noop {} {\bibfield  {journal} {\bibinfo  {journal} {Journal of
  Cosmology and Astroparticle Physics}\ }\textbf {\bibinfo {volume} {2021}},\
  \bibinfo {pages} {055} (\bibinfo {year} {2021})}\BibitemShut {NoStop}%
\bibitem [{\citenamefont {Fluri}\ \emph {et~al.}(2018)\citenamefont {Fluri},
  \citenamefont {Kacprzak}, \citenamefont {Refregier}, \citenamefont {Amara},
  \citenamefont {Lucchi},\ and\ \citenamefont
  {Hofmann}}]{fluri2018cosmological}%
  \BibitemOpen
  \bibfield  {author} {\bibinfo {author} {\bibfnamefont {J.}~\bibnamefont
  {Fluri}}, \bibinfo {author} {\bibfnamefont {T.}~\bibnamefont {Kacprzak}},
  \bibinfo {author} {\bibfnamefont {A.}~\bibnamefont {Refregier}}, \bibinfo
  {author} {\bibfnamefont {A.}~\bibnamefont {Amara}}, \bibinfo {author}
  {\bibfnamefont {A.}~\bibnamefont {Lucchi}}, \ and\ \bibinfo {author}
  {\bibfnamefont {T.}~\bibnamefont {Hofmann}},\ }\href@noop {} {\bibfield
  {journal} {\bibinfo  {journal} {Physical Review D}\ }\textbf {\bibinfo
  {volume} {98}},\ \bibinfo {pages} {123518} (\bibinfo {year}
  {2018})}\BibitemShut {NoStop}%
\bibitem [{\citenamefont {Fluri}\ \emph {et~al.}(2019)\citenamefont {Fluri},
  \citenamefont {Kacprzak}, \citenamefont {Lucchi}, \citenamefont {Refregier},
  \citenamefont {Amara}, \citenamefont {Hofmann},\ and\ \citenamefont
  {Schneider}}]{fluri2019cosmological}%
  \BibitemOpen
  \bibfield  {author} {\bibinfo {author} {\bibfnamefont {J.}~\bibnamefont
  {Fluri}}, \bibinfo {author} {\bibfnamefont {T.}~\bibnamefont {Kacprzak}},
  \bibinfo {author} {\bibfnamefont {A.}~\bibnamefont {Lucchi}}, \bibinfo
  {author} {\bibfnamefont {A.}~\bibnamefont {Refregier}}, \bibinfo {author}
  {\bibfnamefont {A.}~\bibnamefont {Amara}}, \bibinfo {author} {\bibfnamefont
  {T.}~\bibnamefont {Hofmann}}, \ and\ \bibinfo {author} {\bibfnamefont
  {A.}~\bibnamefont {Schneider}},\ }\href@noop {} {\bibfield  {journal}
  {\bibinfo  {journal} {Physical Review D}\ }\textbf {\bibinfo {volume}
  {100}},\ \bibinfo {pages} {063514} (\bibinfo {year} {2019})}\BibitemShut
  {NoStop}%
\bibitem [{\citenamefont {Farsian}\ \emph {et~al.}(2020)\citenamefont
  {Farsian}, \citenamefont {Krachmalnicoff},\ and\ \citenamefont
  {Baccigalupi}}]{farsian2020foreground}%
  \BibitemOpen
  \bibfield  {author} {\bibinfo {author} {\bibfnamefont {F.}~\bibnamefont
  {Farsian}}, \bibinfo {author} {\bibfnamefont {N.}~\bibnamefont
  {Krachmalnicoff}}, \ and\ \bibinfo {author} {\bibfnamefont {C.}~\bibnamefont
  {Baccigalupi}},\ }\href@noop {} {\bibfield  {journal} {\bibinfo  {journal}
  {Journal of Cosmology and Astroparticle Physics}\ }\textbf {\bibinfo {volume}
  {2020}},\ \bibinfo {pages} {017} (\bibinfo {year} {2020})}\BibitemShut
  {NoStop}%
\bibitem [{\citenamefont {Caldeira}\ \emph {et~al.}(2019)\citenamefont
  {Caldeira}, \citenamefont {Wu}, \citenamefont {Nord}, \citenamefont
  {Avestruz}, \citenamefont {Trivedi},\ and\ \citenamefont
  {Story}}]{caldeira2019deepcmb}%
  \BibitemOpen
  \bibfield  {author} {\bibinfo {author} {\bibfnamefont {J.}~\bibnamefont
  {Caldeira}}, \bibinfo {author} {\bibfnamefont {W.~K.}\ \bibnamefont {Wu}},
  \bibinfo {author} {\bibfnamefont {B.}~\bibnamefont {Nord}}, \bibinfo {author}
  {\bibfnamefont {C.}~\bibnamefont {Avestruz}}, \bibinfo {author}
  {\bibfnamefont {S.}~\bibnamefont {Trivedi}}, \ and\ \bibinfo {author}
  {\bibfnamefont {K.~T.}\ \bibnamefont {Story}},\ }\href@noop {} {\bibfield
  {journal} {\bibinfo  {journal} {Astronomy and Computing}\ }\textbf {\bibinfo
  {volume} {28}},\ \bibinfo {pages} {100307} (\bibinfo {year}
  {2019})}\BibitemShut {NoStop}%
\bibitem [{\citenamefont {Pal}\ \emph {et~al.}(2023)\citenamefont {Pal},
  \citenamefont {Chanda},\ and\ \citenamefont {Saha}}]{Pal:2022hpi}%
  \BibitemOpen
  \bibfield  {author} {\bibinfo {author} {\bibfnamefont {S.}~\bibnamefont
  {Pal}}, \bibinfo {author} {\bibfnamefont {P.}~\bibnamefont {Chanda}}, \ and\
  \bibinfo {author} {\bibfnamefont {R.}~\bibnamefont {Saha}},\ }\href {\doibase
  10.3847/1538-4357/acb4ee} {\bibfield  {journal} {\bibinfo  {journal}
  {Astrophys. J.}\ }\textbf {\bibinfo {volume} {945}},\ \bibinfo {pages} {77}
  (\bibinfo {year} {2023})},\ \Eprint {http://arxiv.org/abs/2203.14060}
  {arXiv:2203.14060 [astro-ph.CO]} \BibitemShut {NoStop}%
\bibitem [{\citenamefont {Akrami}\ \emph {et~al.}(2020)\citenamefont {Akrami}
  \emph {et~al.}}]{Planck:2018yye}%
  \BibitemOpen
  \bibfield  {author} {\bibinfo {author} {\bibfnamefont {Y.}~\bibnamefont
  {Akrami}} \emph {et~al.} (\bibinfo {collaboration} {Planck}),\ }\href
  {\doibase 10.1051/0004-6361/201833881} {\bibfield  {journal} {\bibinfo
  {journal} {Astron. Astrophys.}\ }\textbf {\bibinfo {volume} {641}},\ \bibinfo
  {pages} {A4} (\bibinfo {year} {2020})},\ \Eprint
  {http://arxiv.org/abs/1807.06208} {arXiv:1807.06208 [astro-ph.CO]}
  \BibitemShut {NoStop}%
\bibitem [{\citenamefont {Yu}\ and\ \citenamefont
  {Koltun}(2015)}]{yu2015multi}%
  \BibitemOpen
  \bibfield  {author} {\bibinfo {author} {\bibfnamefont {F.}~\bibnamefont
  {Yu}}\ and\ \bibinfo {author} {\bibfnamefont {V.}~\bibnamefont {Koltun}},\
  }\href@noop {} {\bibfield  {journal} {\bibinfo  {journal} {arXiv preprint
  arXiv:1511.07122}\ } (\bibinfo {year} {2015})}\BibitemShut {NoStop}%
\bibitem [{\citenamefont {Hermann}\ \emph {et~al.}(2020)\citenamefont
  {Hermann}, \citenamefont {Chen},\ and\ \citenamefont
  {Kornblith}}]{hermann2020origins}%
  \BibitemOpen
  \bibfield  {author} {\bibinfo {author} {\bibfnamefont {K.}~\bibnamefont
  {Hermann}}, \bibinfo {author} {\bibfnamefont {T.}~\bibnamefont {Chen}}, \
  and\ \bibinfo {author} {\bibfnamefont {S.}~\bibnamefont {Kornblith}},\
  }\href@noop {} {\bibfield  {journal} {\bibinfo  {journal} {Advances in Neural
  Information Processing Systems}\ }\textbf {\bibinfo {volume} {33}},\ \bibinfo
  {pages} {19000} (\bibinfo {year} {2020})}\BibitemShut {NoStop}%
\bibitem [{\citenamefont {Zhang}\ \emph {et~al.}(2019)\citenamefont {Zhang},
  \citenamefont {Wang},\ and\ \citenamefont {Jung}}]{8502129}%
  \BibitemOpen
  \bibfield  {author} {\bibinfo {author} {\bibfnamefont {Z.}~\bibnamefont
  {Zhang}}, \bibinfo {author} {\bibfnamefont {X.}~\bibnamefont {Wang}}, \ and\
  \bibinfo {author} {\bibfnamefont {C.}~\bibnamefont {Jung}},\ }\href {\doibase
  10.1109/TIP.2018.2877483} {\bibfield  {journal} {\bibinfo  {journal} {IEEE
  Transactions on Image Processing}\ }\textbf {\bibinfo {volume} {28}},\
  \bibinfo {pages} {1625} (\bibinfo {year} {2019})}\BibitemShut {NoStop}%
\bibitem [{\citenamefont {Ronneberger}\ \emph {et~al.}(2015)\citenamefont
  {Ronneberger}, \citenamefont {Fischer},\ and\ \citenamefont
  {Brox}}]{ronneberger2015u}%
  \BibitemOpen
  \bibfield  {author} {\bibinfo {author} {\bibfnamefont {O.}~\bibnamefont
  {Ronneberger}}, \bibinfo {author} {\bibfnamefont {P.}~\bibnamefont
  {Fischer}}, \ and\ \bibinfo {author} {\bibfnamefont {T.}~\bibnamefont
  {Brox}},\ }in\ \href@noop {} {\emph {\bibinfo {booktitle} {Medical image
  computing and computer-assisted intervention--MICCAI 2015: 18th international
  conference, Munich, Germany, October 5-9, 2015, proceedings, part III 18}}}\
  (\bibinfo {organization} {Springer},\ \bibinfo {year} {2015})\ pp.\ \bibinfo
  {pages} {234--241}\BibitemShut {NoStop}%
\bibitem [{\citenamefont {Gorski}\ \emph {et~al.}(2005)\citenamefont {Gorski},
  \citenamefont {Hivon}, \citenamefont {Banday}, \citenamefont {Wandelt},
  \citenamefont {Hansen}, \citenamefont {Reinecke},\ and\ \citenamefont
  {Bartelmann}}]{Gorski_2005}%
  \BibitemOpen
  \bibfield  {author} {\bibinfo {author} {\bibfnamefont {K.~M.}\ \bibnamefont
  {Gorski}}, \bibinfo {author} {\bibfnamefont {E.}~\bibnamefont {Hivon}},
  \bibinfo {author} {\bibfnamefont {A.~J.}\ \bibnamefont {Banday}}, \bibinfo
  {author} {\bibfnamefont {B.~D.}\ \bibnamefont {Wandelt}}, \bibinfo {author}
  {\bibfnamefont {F.~K.}\ \bibnamefont {Hansen}}, \bibinfo {author}
  {\bibfnamefont {M.}~\bibnamefont {Reinecke}}, \ and\ \bibinfo {author}
  {\bibfnamefont {M.}~\bibnamefont {Bartelmann}},\ }\href {\doibase
  10.1086/427976} {\bibfield  {journal} {\bibinfo  {journal} {The Astrophysical
  Journal}\ }\textbf {\bibinfo {volume} {622}},\ \bibinfo {pages} {759–771}
  (\bibinfo {year} {2005})}\BibitemShut {NoStop}%
\bibitem [{\citenamefont {Wang}\ \emph {et~al.}(2004)\citenamefont {Wang},
  \citenamefont {Bovik}, \citenamefont {Sheikh},\ and\ \citenamefont
  {Simoncelli}}]{1284395}%
  \BibitemOpen
  \bibfield  {author} {\bibinfo {author} {\bibfnamefont {Z.}~\bibnamefont
  {Wang}}, \bibinfo {author} {\bibfnamefont {A.}~\bibnamefont {Bovik}},
  \bibinfo {author} {\bibfnamefont {H.}~\bibnamefont {Sheikh}}, \ and\ \bibinfo
  {author} {\bibfnamefont {E.}~\bibnamefont {Simoncelli}},\ }\href {\doibase
  10.1109/TIP.2003.819861} {\bibfield  {journal} {\bibinfo  {journal} {IEEE
  Transactions on Image Processing}\ }\textbf {\bibinfo {volume} {13}},\
  \bibinfo {pages} {600} (\bibinfo {year} {2004})}\BibitemShut {NoStop}%
\bibitem [{\citenamefont {Lewis}\ \emph {et~al.}(2000)\citenamefont {Lewis},
  \citenamefont {Challinor},\ and\ \citenamefont
  {Lasenby}}]{lewis2000efficient}%
  \BibitemOpen
  \bibfield  {author} {\bibinfo {author} {\bibfnamefont {A.}~\bibnamefont
  {Lewis}}, \bibinfo {author} {\bibfnamefont {A.}~\bibnamefont {Challinor}}, \
  and\ \bibinfo {author} {\bibfnamefont {A.}~\bibnamefont {Lasenby}},\
  }\href@noop {} {\bibfield  {journal} {\bibinfo  {journal} {The Astrophysical
  Journal}\ }\textbf {\bibinfo {volume} {538}},\ \bibinfo {pages} {473}
  (\bibinfo {year} {2000})}\BibitemShut {NoStop}%
\bibitem [{\citenamefont {Kingma}(2014)}]{kingma2014adam}%
  \BibitemOpen
  \bibfield  {author} {\bibinfo {author} {\bibfnamefont {D.}~\bibnamefont
  {Kingma}},\ }\href@noop {} {\bibfield  {journal} {\bibinfo  {journal} {arXiv
  preprint arXiv:1412.6980}\ } (\bibinfo {year} {2014})}\BibitemShut {NoStop}%
\end{thebibliography}%

\vfill ${}$

\end{document}